\documentclass[modern]{aastex631}
\usepackage{amsmath,array,graphicx}
\usepackage{tabularx}
\usepackage{float}

\def\lapprox{\hbox{\lower .8ex\hbox{$\,\buildrel < \over\sim\,$}}}
\def\gapprox{\hbox{\lower .8ex\hboxeening{$\,\buildrel > \over\sim\,$}}}

\def\aap{\hbox{A\&A}}

\begin{document}

 \title
    {''SNe Ia twins for life'' towards 
     a precise determination of H$_{0}$}

\bigskip

\author{P. Ruiz-Lapuente}
\email{pilar@icc.ub.edu}
\affiliation{Instituto de F\'{\i}sica Fundamental, Consejo Superior de 
Investigaciones Cient\'{\i}ficas, c/. Serrano 121, E-28006, Madrid, Spain}
\affiliation{Institut de Ci\`encies del Cosmos (UB--IEEC),  c/. Mart\'{\i}
i Franqu\'es 1, E--08028, Barcelona, Spain}

\bigskip

\author{J.I. Gonz\'alez Hern\'andez}
\affiliation{Instituto de Astrof\'{\i}sica de Canarias, E-38200 La Laguna, 
Tenerife, Spain}
\affiliation{Universidad de La Laguna, Dept. Astrof\'{\i}sica, E38206 
La Laguna, Tenerife, Spain}

\bigskip

\begin{abstract}

   Here we present an approach to the measurement of extragalactic distances
   using twin SNe Ia, taken from the early down to the nebular phase.
   The approach is purely empirical, although we can give a
   theoretical background on why the method is reliable.
    By studying those twins in galaxies where peculiar velocities
   are  relatively unimportant, we can tackle the 
   H$_{0}$ tension problem. Here we apply the method to the determination of the
   distances to NGC 7250 and NGC 2525, who hosted respectively SN 2013dy and
   SN 2018gv, twins of two different SNe Ia prototypes: SN 2013aa/
   SN 2017cbv and SN2011fe. 
   From the study of the SN 2013aa and SN 2017cbv twin pair, by comparing it
   with
   SN 2011fe and applying the difference between the SN 2013aa/2017cbv and 
   the SN 2011fe class, we find as well a good estimate of
   the distance to NGC 5643.  
   We have just started to measure distances with this method for
   the samples in Freedman et al. and Riess et al. There
   are differences in measured distances to the same galaxy using Cepheids or
   TRGBs. In this context of discrepancy, the ''twins for life '' method
   is very competitive because it can provide distance estimates  with a modulus
   error of $\sigma_{\mu}$ $=$ 0.04 mag. Our findings called for a revision of
   the distances measured with Cepheids in Riess et al. (2022).
   NGC 7250 and NGC 2525 needed better measurements with Cepheids. We
   have noticed that the Cepheids--based distance obtained with the {\it JWST}
   in Riess et al. (2024a) for NGC 5643 is in a a good agreement
   with what we find, unlike their previous estimate in Riess et al. (2022).
   The Hubble tension can arise from the way in which the local SNe Ia
   sample is linked to the SNe Ia  Hubble flow sample. A good calibration
   of SNe Ia in the local sample is needed and we have started to gather it.  
   We also expect to apply the ''twin'' SNe Ia comparison from the local sample
   to that in galaxies with z $>$0.02--0.03
   well into
   the Hubble flow to obtain a reliable value for H$_{0}$.  Those distant
   SNe Ia can be observed with 
   the {\it ELT} or the
   {\it JWST}. 
   
 \end{abstract}


\section{Introduction}

\bigskip

\noindent
Type Ia supernovae (SNe Ia) at high $z$ led to the discovery of the acceleration
of the Universe (Riess et al. 1998; Perlmutter et al. 1999) and of the
presence of a repulsive dynamical component that permeates it and accounts for
such accelerated expansion. This component is known as dark energy and its
nature is the subject of present and future research.

\bigskip

\noindent
Apart from such endeavour in the field of very large
$z$ (see the latest developments in Rubin et al. 2023),
there is, at low $z$ a very important question to solve: the Hubble tension.
This is a well known  tension between the value of the Hubble
constant H$_{0}$ derived from the CMB 
by the {\it Planck} collaboration and that obtained, using Cepheids, by Riess
  and co-workers in their {\it SH0ES} program.  The CMB value of H$_{0}$ =
  67.4$\pm$0.5 km s$^{-1}$ Mpc$^{-1}$ (Planck Collaboration 2020) and the latest
 {\it SH0ES}
  value of  H$_{0}$ = 73.29$\pm$0.9 km s$^{-1}$ Mpc$^{-1}$ (Murakami et al et. 2023) have now a discrepancy at the 5.7 $\sigma$ level.

  \bigskip

  \noindent
  Such 
difference in H$_{0}$ questions seriously the
$\Lambda$CDM  model and some authors claim  
the need of new physics in the early Universe to make the
{\it Planck} value compatible with that derived by methods
involving low-$z$ astrophysical distance indicators such as Cepheids
(see Di Valentino
et al. 2021 for a review). An exploration for a possible  consistency of
the CMB Planck  data with the  {\it SH0ES} H$_{0}$ has been done
(see the early attempt by  Bernal et al. 2016,  for instance),
but  no consistent solution appears satisfactory (Efstathiou et al. 2023,
amongst others). 

\bigskip

\noindent
The situation, however, is not completely clear, since there are methods that
do not
use Cepheids and  predict
values of H$_{0}$ in-between those from {\it Planck} and from
{\it SH0ES}. 
So, using the {\it Tip of the Red Giant Branch
(TRGB)} method, Freedman et al. (2019)  found a value of H$_{0}$ =
69.8$\pm$0.8 (stat)$\pm$1.7  (sys) km  s$^{-1}$ Mpc$^{-1}$.
They noted a mean difference
in galaxy moduli $\mu^{TRGB} - \mu^{Ceph}$ = 0.059 mag. They measured a 
scatter significantly larger for the distant galaxies
than for the nearby ones, where it amounts to $\pm$
0.05 mag only.  More recently, Madore \& Freedman (2024) note that
  the more--distant sample of galaxies has both larger scatter in
  the sample and systematically lower Cepheid distance moduli than
  their corresponding TRGB distance moduli. In the TRGB versus Cepheids
  comparison differences,  one can see
  an average error in the differences between  moduli for the near and the
  far sample (lower or above 12.5 Mpc) of 0.100 and 0.103 mag, respectively, 
  and a scatter of the data points around the mean of
  0.068 and 0.152 mag, respectively. This analysis
  suggests that the errors on the nearby
  galaxies moduli are overestimated by 0.032 mag and the errors on the
  distant sample unerestimated by 0.049 mag. Thus, the discrepancies need a
  deeper examination between the two methods. In the meanwhile,  Anand et al.
  (2024) have examined the difference in distance moduli to two
  galaxies using {\it JWST} data and applying  both the
  TRGB and the Cepheids
  approaches and find a better agreement between those distances.
  This path is worth to be extended
  to the 37  SNe Ia host galaxies used in distance determinations by
   Riess et al. (2022). Most recently, there has been a release
     of the results from {\it JWST} Cepheids, {\it JWST} TRGB stars and
     {\it JWST} JAGB (J--region Asymptotic Giant Branch) stars by 
     Freedman et al. (2024) and a reevaluation by
     Riess et al. (2024b). The number of
     distances to galaxies obtained with the {\it JWST} is limited to 11. 
     There is a better agreement between distances from both
     teams, but the H$_{0}$ obtained by each of the methods differ
     substantially. One major surprise has been the value obtained
     by Freedman et al. (2024) with JAGBs, which basically suggests
     that there is no Hubble Tension. The three methods give different
     values for H$_{0}$ in Freedman et al. (2024):
     from TRGB H$_{0}$ $=$ 69.85 $\pm$ 1.75 (stat) $\pm$ 1.54 (sys),
     from Cepheids H$_{0}$ $=$ 72.05 $\pm$ 1.86 (stat) $\pm$ 3.10 (sys),
     from JAGBs  H$_{0}$ $=$ 67.96 $\pm$ 1.85 (stat) $\pm$ 1.90 (sys).
     In this study the changes in H$_{0}$ can be traced to different
     distance moduli from various methods. As mentioned in
     Freedman et al. (2024), the weighted (unweighted) mean
     difference  between JAGB minus Cepheid distance moduli is 0.086 $\pm$ 0.028
     (0.083 $\pm$ 0.031) mag or 4 $\%$. That can drive the change in
     the value of H$_{0}$. Though, the distances by Riess et al. (2022, 2024a,b)
     are now significantly closer to the distances obtained
     by {\it JWST} by Freedman et al. (2024), Riess et al (2024b)
     get very different numbers
     in their final H$_{0}$, those being for
     {\it JWST} Cepheids 73.4 $\pm$ 2.1, for {\it JWST} JAGBs
     is 72.2 $\pm$ 2.2 and
     for {\it JWST} TRGBs 72.1 $\pm$ 2.2 km s$^{-1}$ Mpc$^{-1}$. The issue
     of why the final value by the two different collaborations differs
     has not been clarified yet. The differences arising
     from the various methods
     require a deeper research.

\bigskip

\noindent
On the other hand, the sample of distances obtained by the {\it Surface
Brightness Fluctuations (SBF)} method, dominated by early-type galaxies as 
required by the method, show a mean difference of 
$\mu^{SFB} - \mu^{Ceph}$ = 0.07 mag between the distances estimated using SFB
and those using the Cepheid calibration (Khetan et al. 2021). A value of
H$_{0}$ =70.50 $\pm$ 2.37 (stat) $\pm$ 3.38 (sys)  km s$^{-1}$ Mpc$^{-1}$
is derived in that paper. This, however, has been questioned by
Blakeslee et al. (2021), obtaning an H$_{0}$ around
73 km s$^{-1}$ Mpc$^{-1}$.

\bigskip

\noindent
In general, there are promising ideas to give an estimate of 
H$_{0}$, but different applications of the methods lead  also to different
results (see the review by Freedman \& Madore 2023, for instance).

\bigskip

\noindent
 To determine the value of H$_{0}$ from the distances to galaxies which are
not yet in the Hubble flow, peculiar velocities must be
considered. 
Kenworthy et al. (2022), studying the uncertanties in the peculiar
velocities obtained by various models, bring  H$_{0}$ from the use of
Cepheids in 35 extragalactic hosts to a range from 71.8 to 77 km  s$^{-1}$ Mpc$^{-1}$. So the actual {\it SH0ES} value for  H$_{0}$ becomes 73.1 $^{+2.6}_{-2.3}$  km
s$^{-1}$ Mpc$^{-1}$, at 2.6$\sigma$ tension with Planck. 
This is a two--rung  distance ladder result which has considered previous
small--scale galactic motions studies from Peterson et al. (2022) and additional
corrections from large--scale structure. 

\bigskip

\noindent
 The measurement of H$_{0}$ with the 
  Tip of the Red Giant Branch and Cepheids  is centered in providing solid
distances to galaxies at $<$ 40 Mpc and  use another brighter distance indicator to reach the Hubble flow. With the advent of the
{\it JWST}, it has been considered the possibility of reaching the Hubble
flow (z$\sim$ 0.02-0.03) with TRGB/Cepheids or with the
relatively new method  J-Region Assymptotic  Giant Branch (JAGB) that 
provides an intrinsically brighter standard candle. The data to go through
this innovative path are still to be gathered.
 Up to the present, the most common way to obtain  H$_{0}$ using these
  methods is to establish  TRGB, Cepheids or JAGB stars
  distances to Type Ia supernovae (SNIa) host galaxies in the local volume
  and then
calibrate together those SNe Ia absolute magnitudes with
SNe Ia in the Hubble flow. The SNe Ia calibration depends therefore
on three rungs:
Rung 1 is the calibration of TRGB/Cepheids (JAGBs)  with local anchors, rung 2
refers to the calibration of the distance to SNe Ia host galaxies with
the TRGB/Cepheids/JAGBs and rung 3 is the  overall calibration of SNe Ia in the
Hubble flow, which is done in different ways but usually
involves the derivation of their absolute magnitudes in relation to H$_{0}$
(see for instance Hamuy et al. 2021; Freedman et al. 2019). 
Hamuy et al. (2021) have found that the final H$_{0}$ relies very much on the
values of the distances to nearby host galaxies of SNe Ia using either the TRGB
or the Cepheids method, i.e, on the second rung. 

\bigskip

\noindent
 Here we address the distances to SNe Ia which constitute the basis for
the H$_{0}$  calibration in rung 2, i.e. in
nearby galaxies that hosted  SNe Ia.
We compare them with the TRGB and Cepheids distances in those galaxies.  We
use the twinness of SNe Ia (see below) and hope, in the future, to reach H$_{0}$  by 
directly allowing twin to twin SNe Ia comparison from nearby z up to
z $\sim$ 0.02-0.03. 
\bigskip

\noindent
The use of SNe Ia as distance indicators for cosmology was only made possible
when the correlation between
maximum brightness and rate of decline in the light curve was formulated in a
general, quantitative way by
Phillips (1993). Later on, further ways to formulate such correlation have
been the stretch $s$, first introduced by Perlmutter et al. (1999), the
{\it Multi Light Curves Shape (MLCSk2)} (Riess et al. 1996;
  Jha et al. 2007), or  
  the stretch $x1$ (Guy et al. 2005), which go beyond the parameterization by
  $\Delta m_{15}$ (magnitude different between peak brightness and that 15 days
  later) of the decline rate introduced by  Phillips (1993).
One current advance in achieving a more precise use of the rate of decline of
the SNe Ia to standardize them
is the inclusion of twin embedding (Fakhouri et al. 2015; Boone et al.
2016, 2021). It started when Fakhouri et al. (2015) found that by
using SN Ia pairs with closely matching spectra, from the SNFactory sample,
they achieved a reduced dispersion in brightness. They were able to
standardize  SNe Ia in the redshift range z between 0.03 and 0.08
to within 0.06–0.07 mag. The idea is
most promising, given the expected massive gathering of SNe Ia spectra at
high $z$. Twins are used there to connect
the absolute magnitude estimate at maximum brightness
with parameters corresponding to the
spectral diversity of the SNe Ia. The method is as well
suited for the time when dark energy is entering in the Hubble diagram,
allowing to determine $\Omega_{M}$ and $\Omega_{DE}$, as well as the
equation of state of dark energy. 

\bigskip

\noindent
In the present paper
we propose to use ''SNe Ia twins for life'', from early to the nebular phases\footnote{For those readers that are not familiar with SNe Ia phases, we clarify here that early phases are those next to maximum brightness, i.e, from -17 days to
  +15 days around the peak of brightness; nebular phases are those typically corresponding to more than 200 days after maximum, and the phase between maximum until the nebular phase corresponds to the evolution from early to late times, often  simply called postmaximum or denoted by the epoch.}.
It is a use of  SNe Ia  to address the Hubble tension issue which uses
the spectral evolution of those SNe Ia which have not only similar stretch but
identical features along the lifetime of the SNe Ia. The precision of the
distance provided by those twins should be larger than for  SNe Ia with just
 similar stretch.

\bigskip

\noindent
The method does not
start from the  apparent magnitudes of the SNe Ia, but from their spectra,
with particular emphasis on those from the nebular phase, which are very
sensitive to the reddening and to the intrinsic luminosity of the SNe. 
To get a direct check of the distance ladder, we make an empirical use 
of twin SNe Ia by comparing their spectra all the way from the early until the
nebular phase. We obtain a good estimate of the relative distances
between SNe Ia which are in different galaxies but seem identical
spectroscopically, in addition to having a similar stretch.  It
is important, here, that the SNe Ia do show twin behaviour through
the SN Ia life.
Given the range of diversity of the SNe Ia already sampled, it should always
be possible to find some of them with spectra similar to those of any newly
discovered SN Ia. 
As nearby SNe Ia will be observed by the 
thousands
per month with the advent of more SNe Ia dedicated surveys with new
telescopes together with the present ones, it will be possibly to match
their spectra with those of 
distant SNe Ia and then gain a new perspective on the extragalactic
distance scale.
Extending this procedure to reach SNe Ia that are in the Hubble flow should
allow to avoid using a fiducial absolute magnitude M$_{B}$--H$_{0}$ relation, and get a direct
comparison of distances  leading straightforwardly to the value of H$_{0}$.
This would be very useful in helping to solve the Hubble tension problem.

\bigskip

\noindent
The gain of using the ''twins for life''  approach is that it provides a
direct measurement of distance, intrinsic color and reddening caused by
Galactic and extragalactic dust by the use of the whole spectra of the SN Ia.
It allows the consistent pairing of SNe Ia through all phases. The selection of 
twins is made of SNe Ia with a similar stretch, being then of similar luminosities, but in addition the ''twinness factor'' can make more precise the distance
estimate with  a  modulus error of 0.04 mag in all filters, as we will show. So, all this makes a very useful tool to gather the right distance
ladder. 

\bigskip

\noindent
     Twins allow to determine distances to SNe Ia host galaxies both
      in relation one to another  in a relative scale  and
     in absolute  scale using well calibrated distances from 
     recent nearby normal SNe Ia like SN 2011fe in M101.
      At present, there is no coincidence on distances to a nearby
       galaxy like M101, but there is progress in that direction. One can, in
       the meanwhile, aim towards a first step in the distance
       ladder by a choice of concordance between methods as  Cepheids with
       {\it JWST}
    data, Miras distances, TRGB, JAGB stars, knowledge from constraints
    on very nearby SNe Ia or through a choice of some other
    method. We will address
    this in some subsections. At low distances like that of M101, the
    difference between methods is much smaller than at much larger distances
    where some of those methods
    can only get an accurate result with {\it JWST} data.

\bigskip

\noindent
In section 2, we describe the method and characterize what is a perfect twin
against other ways to compare SNe Ia. In section 3, we  obtain
the distance to NGC 7250, the host galaxy of SN 2013dy. In section 4,
we  obtain
the distance to NGC NGC 2525, the host galaxy of SN 2018gv. In section 5
we
compare our results within the SNe Ia diversity and discuss them. 
In section 6  we give our conclusions.

\section{Purely empirical twins until the nebular phase }

\bigskip

\noindent
The key to this approach is to have a nearby SN Ia of the same type as
another at much longer distance. For the nearby SN Ia, a good distance
determination should be available. We will demonstrate this by using
SN 2013dy in NGC 7250, which is a twin of SN2017cbv/SN2013aa both in NGC 5643.

\bigskip

\noindent
Table 1 show the references for the spectra and photometry used in this paper.
Some spectra have been provided generously by various researchers,
and some others have been taken from  the {\it WISeREP}. All are
carefully calibrated in flux in agreement with the published photometry.
The error in the flux calibration is well taken into account. 

\bigskip

{\bf

\begin{table}[h!]
  \centering
  \caption{Spectra and photometry of the SNe Ia twins sample}
  \begin{tabular}{llcrll}
    \hline
    \hline
    {\bf Spectra} &  &   &     &     &      \\
    \hline
    SN & Galaxy & MJD & Phase (days) & Reference$^{*}$ & Comments \\
    \hline
    2013aa & NGC 5643 & 56341 &  -2  & Burns et al. (2020) & Priv. comm. \\
    $\dots$ & $\dots$ & 56391 & +48 &         *      &   \\
    $\dots$ & $\dots$ & 56676 & +333 &        *       &    \\
    $\dots$ & $\dots$ & 56704 & +361 &        *       &       \\
    2017cbv & NGC 5643 & 57836 & -2   & Burns et al. (2020) & Priv. comm. \\
    $\dots$ & $\dots$ & 57848 & +12   &       *       &             \\
    $\dots$ & $\dots$ & 58171 & +333 &        *    &      \\
    $\dots$ & $\dots$ & 58199 & +361 &        *        &      \\
    2013dy & NGC 7250 & 56498 &  -2  & Pan et al. (2015) &     \\
    $\dots$ & $\dots$ & 56546 & +46  & Zhai et al. (2016) &    \\
    $\dots$ & $\dots$ & 56833 & +333 & Zhai et al. (2016) &   \\
    2011fe & M101     & 55823 &  +9  & Zhang et al. (2016) &  \\
    $\dots$ & $\dots$ & 55826 & +12  &        *            &  \\
    $\dots$ & $\dots$ & 56103 & +289 & Mazzali et al. (2015) &  \\
    $\dots$ & $\dots$ & 56158 & +344 &        *       &       \\
    2018gv & NGC 2525 & 58159 &  +9  &        *         &  \\
    $\dots$ & $\dots$ & 58439 & +289  & Graham et al. (2022) & Priv. comm. \\
    $\dots$ & $\dots$ & 58494 & +344  & Graham et al. (2022) & Priv. comm. \\
    \hline
    {\bf Photometry} &  &   &      &     &    \\
    \hline
    2013aa & --- &  ---    &  ---    & Burns et al. (2020)     & --- \\
    2017cbv & --- &  ---    &  ---    & Burns et al. (2020)    & --- \\
    2013dy & --- & --- & --- & Pan et al. (2015) & --- \\
    2011fe & --- & ---  & --- & Zhang et al. (2016) & --- \\
    2018gv & --- &---  & --- & Scolnic et al. (2022) & --- \\
    \hline  
  \noindent{*  Spectrum from the {\it WISeREP}}.

\end{tabular}
\end{table}

}

\begin{table}[ht!]
  \centering
  \caption{SN 2013aa and SN 2017cbv}
  \begin{tabular}{lcc}
    \hline
    \hline
        {\bf SN 2013aa} &      &     \\
        RA, DEC$^{a}$ &  14:32:33.881 & -44:13:27.80 \\
        Discovery date$^{a}$ & $\dots$ &  2013-02-13 \\
        Phase ( referred to maximum light)$^{b}$& $\dots$ & -7 days \\
        Redshift$^{c}$ & $\dots$ & 0.004 \\ 
        E(B-V)$_{MW}$$^{d}$ & $\dots$ & 0.15$\pm$0.06 mag \\
        $m_{B}^{max}$ $^{b}$ & $\dots$ & 11.094$\pm$0.003 mag \\
        $\Delta m(B)_{15}$$^{b}$ & $\dots$ & 0.95$\pm$0.01 \\
        Stretch factor $s_{BV}^{D}$$^{b}$ & $\dots$ & 1.11$\pm$0.02 \\
        Phases of the spectra used & $\dots$ & -2, 360-370 days \\
        \hline 
            {\bf SN 2017cbv} &     &    \\
            RA, DEC$^{e}$ & 14:32:34.420 & -44:08:02.74 \\
            Discovery date$^{e}$ & $\dots$ & 2017-03-10 \\
            Phase (referred to maximum light)$^{b}$ & $\dots$ & -18 days \\
            Redshift$^{c}$ & $\dots$ & 0.004 \\
            E(B-V)$_{MW}$$^{d}$ & $\dots$ &  0.15$\pm$0.06 mag \\
            $m_{B}^{max}$ $^{b}$ & $\dots$ & 11.11$\pm$0.03 mag \\
            $\Delta m(B)_{15}$$^{b}$ & $\dots$ & 0.96$\pm$0.02 \\
            Stretch factor $s_{BV}^{D}$$^{b}$ & $\dots$
            & 1.11$\pm$0.03 \\
            Phases of the spectra used & $\dots$ & -2, 360-370 days \\
            \hline
  \end{tabular}
   \begin{tabular}{lll}    
    $^{a}$Waagen (2013). & $^{b}$Burns et al. (2020). & $^{c}$Parrent et al.
    (2013). \\
    $^{d}$Schlafly \& Finkbeiner (2011). & $^{e}$Tartaglia et al.
    (2017). &  \\
 \end{tabular}
\end{table}

\noindent
We first check the method by taking two twin SNe Ia  
that are at the same distance and share the
same reddening: SN 2017cbv/SN 2013aa.

\subsection{Testing the method with twin SNe Ia in NGC 5643}

\bigskip

\noindent
SN 2013aa and SN 2017cbv appeared in the same galaxy, NGC 5643. They had
similar decline rates and similar B-peak magnitudes.
The studies done on
these two SNe Ia also reveal similar characteristics in other aspects. Table
1 specifies the decline
rates of those SNe Ia not only by $\Delta m_{15} (B)$  but also by s$_{BV}^{D}$
(Burns et al. 2020).

\bigskip

\noindent
Burns et al. (2014) showed that the
color-stretch parameter is a robust way to classify
SNe Ia in terms of light-curve shape and  intrinsic colors.
They propose a  way to quantify the decline rate of the SNe
directly from photometry, by just determining the epoch when the
B--V color curves reach their maxima (i.e., when the SNe
are reddest) relative to that of the
B-band maximum. Dividing this time interval by 30 days then gives the observed
color-stretch parameter s$_{BV}^{D}$. the ``D'' in the superindex
refering to the direct comparison in B, without having to take into account the
BVRI light curves.

\bigskip

\noindent
The problems
involved in the use of $\Delta m_{15} (B)$ are known. The first one is that the decline in 15 days past
maximum can take (and it often does) differing values in different authors.
Phillips et al. (1999), noted in addition that 
any reddening undergone
by the SN would change the shape of the B-band light-curve and
therefore the observed value of $\Delta m(B)_{15}$. A similar problem
is that, by definition, $\Delta m(B)_{15}$ is tied to a particular
photometric system, and so will vary from data set to data set and
will require S-corrections (Suntzeff et al. 1988; Stritzinger
et al. 2002) to convert one set of $\Delta m_{15} (B)$ to another. And
lastly, $\Delta m(B)_{15}$ is defined by measuring the light curve at two
very specific epochs, and some form of interpolation is usually needed
to obtain these values.

\bigskip

\noindent
Fitting the B-V
color curves with cubic splines, Burns et al (2020)
ﬁnd identical color-stretch s$_{BV}^{D}$ = 1.11 for SN 2013aa and SN 2017cbv.
This value is smaller 
 (slower decline) than in typical SNe Ia. They also found similar values in the
 {\it Branch classification}, concerning the equivalent widths (EW) of
 the lines of different elements. This proves that their interior was
 very alike in chemical composition. 

\bigskip

\noindent
SN 2013aa and SN 2017cbv present a unique
opportunity to test the method, by checking whether it gives any
difference in
the distance moduli of these two SNe Ia, using
early and nebular spectra together. Both happened in
NGC 5643. They were in the outskirts of the host galaxy, so the reddening
E(B-V) there should be negligible. The reddening
in our Galaxy, in the direction of NGC 5643, is $(E-B)_{MW}$ = 0.15 mag, 
from Schlafly \& Finkbeiner (2011). The properties of these two SNe Ia can
be found in Table 2.

\bigskip

\subsection{Methodology}

\bigskip

\noindent
The method to determine the relative distance $\Delta \mu$ between the two
SNe Ia will also give the relative reddening $\Delta E(B - V)$.
  Its value will indicate whether $E(B–V)$ is
  larger or smaller for one of the two SNe with respect to its twin, 
  taken as reference. The result
  in $\Delta \mu$ indicates the difference in distance between that SN and
  the reference. The twin of reference is ideally a SN Ia for which
  the distance is well known. It has the role of anchor.

\bigskip

\noindent
In the present case, we take as reference any of the twins. 
  We do not need to know its distance: the result should point
  to a $\Delta \mu$ $\sim$ 0 anyway.
  We will compare SN 2013aa with respect to the class represented by
  SN 2017cbv. The two
  phases taken into account are -2 days before and
  361 days after B-maximum.  Both SNe Ia happened in the outskirts
    of the same galaxy, NGC 5643, thus having negligible reddening in
    the host galaxy.

\bigskip

\noindent
To find the best values and the uncertainties of all variables, we explore the
parameter space with Markov Chain Monte Carlo (MCMC) techniques after
converting the $\chi^{2}$ into a log likelihood function:

\begin{equation}
{\rm log} \mathcal{L} = -0.5 * \sum{(f_{\lambda}^{obs} -
  f_{\lambda}^{ref})^{2}/\sigma_{\lambda}^{2}} + {\rm log} \sigma_{\lambda}
\end{equation}

\bigskip

\noindent
where {\it obs} corresponds to a given supernova and {\it ref} to its  reference
twin, which represents
a whole class. $f_{\lambda}^{obs}$ and $f_{\lambda}^{ref}$
correspond to the fluxes of the two SNe at different wavelengths, and
$\sigma_{\lambda}$
includes the quadratic addition of the uncertainty on the fluxes of both SNe.

\bigskip

\noindent
In fact, as we are using different phases, we have a total likelihood function.
For two phases, $n$ corresponds to 1 and 2 below. Thus the total likelihood
function is written as

\begin{equation}
{\rm log} \mathcal{L}_{n} = {\rm log} \mathcal{L}_{1} +
{\rm log} \mathcal{L}_{2}
\end{equation}

\bigskip
  
  \noindent
  i.e. just the addition of all individual SN phases.

\bigskip

\noindent
We use a Python package, EMCEE (Foreman–Mackey et al. 2013)\footnote{https://emcee.readthedocs.io/} to explore the
likelihood of each variable. EMCEE utilizes an affine invariant MCMC ensemble
sampler proposed by Goodman \& Weare (2010). This sampler
tunes only two parameters to get the desired output: number of
walkers and number of steps.
The run starts by assigning initial maximum likelihood values of the
variables to the walkers. The walkers then start wandering and explore the
full posterior distribution.
After an initial run, we inspect the samplers for their performance. We
do this by looking at the time series of variables in the chain and computing
the autocorrelation time, $\tau$\footnote{https://emcee.readthedocs.io
/en/stable/tutorials/autocorr/}. In our case, the maximum autocorrelation
time among the different variables is about 50.
 When the chains are sufficiently burnt-in
(e.g., they forget their initial start point), we can safely throw away some
 steps that are a few times higher than the burnt-in steps. In our case,
 we run EMCEE with 32 walkers and 10,000 steps and throw
away the first  $\sim$ 250 samples, equivalent to $\sim$ 4 times
the maximum autocorrelation time. 
Thus, our burn--in value in this computation is equal to 4 times the maximum autocorrelation time. 
  A criterion
of good sampling is the acceptance fraction, $a_{f}$. This is the fraction of
steps that are accepted after the sampling is done. The suggested value of
$a_{f}$ is between 0.2 - 0.5 (Gelman et al. 1996).
In each run, we typically  obtained $a_{f} \sim$ 0.25.

\bigskip

\noindent
We adopt uniform priors on the distance $\Delta \mu(\mathcal{U}[-0.3,0.3]$) mag and
reddening $\Delta E(B - V) (\mathcal{U}[0.0, 0.3]$ )mag.

\bigskip

\noindent
One can visualize the output of
two-dimensional and one-dimensional posterior probability distributions in a
corner plot corresponding to 1$\sigma$, 2$\sigma$, 3$\sigma$. The
results point to $\Delta E(B - V)$ = 0.0$\pm$0.0, as it should have been
expected from the two SNe being in the same galaxy  with similar
 conditions of negligible reddening in the host galaxy
  and same reddening by dust in our Galaxy, and a $\Delta\mu$ of
0.004 $\pm$ 0.005 in the early spectrum and $\mu$ =--0.023$^{+0.008}_{-0.007}$
in the nebular one. Those correspond
to a precision of 23 kpc for the early spectrum and around 100 kpc for the late
one. In this plot, both the early and the late-time spectra are used.
It is expected that individual fits will show their own errors.  
Separate comparisons for the two different phases are shown in Figure 1
and Figure 2.
A joint comparison is seen in Figure 3. 
 The joint comparison allows to see what happens if the various epochs
  are fitted together. $\Delta\mu$ $=$--0.005 $\pm$ 0.004, if we fit together the two phases.

\bigskip

\noindent
The preceding demonstrates very successfully the possibility to combine early
and late phase information to obtain relative distances or differences in
distance moduli between the hosts of twin SNe Ia. 
 
\bigskip

\noindent
SN 2013aa and SN 2017cbv are their own SN Ia subclass. SN 2013dy belongs
to that class of twinness. 

\bigskip

\noindent
     In order to estimate the magnitude accuracy that twins have
      in their spectral
  comparison, we calculate the difference in the various filters B, V,R, I
 of the twin spectra at the same phase once they are shifted to a same distance
 (this same distance is calculated given the information provided by our emcee
 computation). The error estimate is of  $\sigma$ $=$ 0.04 in all filters
 and all phases. We have done the comparison not only with SN 2017cbv and
 SN 2013aa, but also with SN 2013dy. In all cases, the above $\sigma$ is
 found. This confirms the power of this method to disentangle
 errors in SNe Ia host galaxies distances provided by TRGB, Cepheids, JAGB. The
 ''twins for life'' method is very competitive.

\bigskip

\begin{figure}[H]
  \centering
\includegraphics[width=0.53\textwidth]{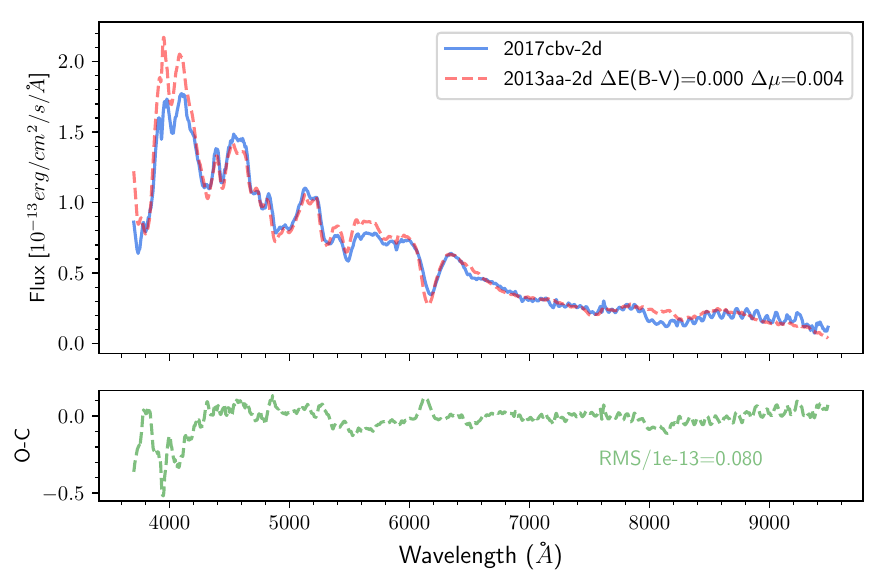}
\includegraphics[width=0.42\textwidth]{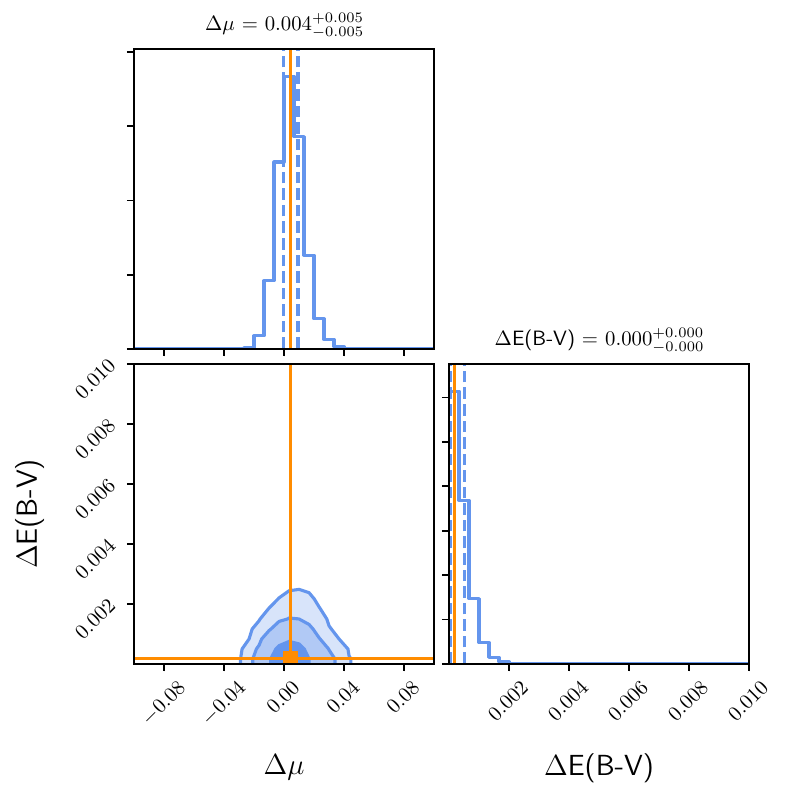}
\caption{Left: Comparison of early time spectra of SN 2013aa and SN 2017cbv at
  -2 days before maximum light. Right:  1$\sigma$, 2$\sigma$ and 3$\sigma$
  contours for the posterior probability distribution. The dashed vertical values correspond to 1$\sigma$ error in each variable.}
  \end{figure}

\begin{figure}[H]
\includegraphics[width=0.53\textwidth]{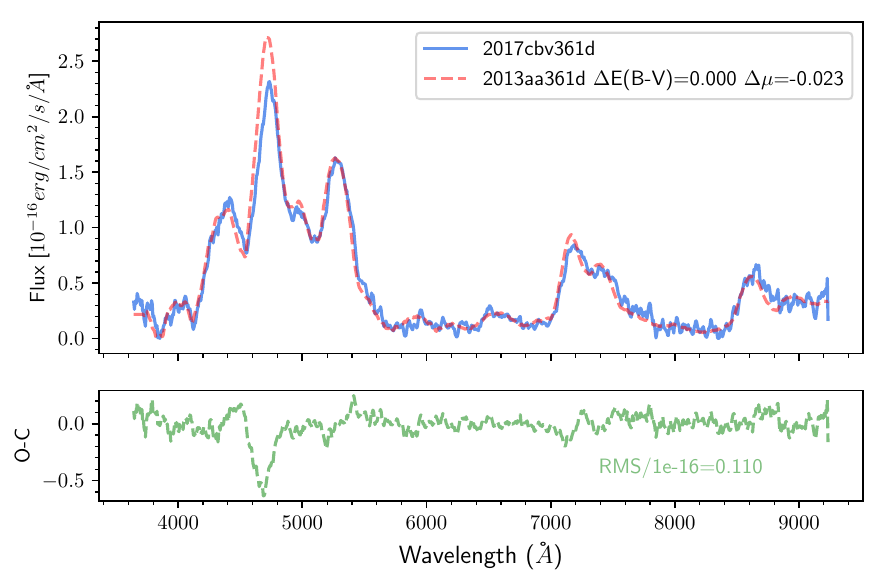}
\includegraphics[width=0.42\textwidth]{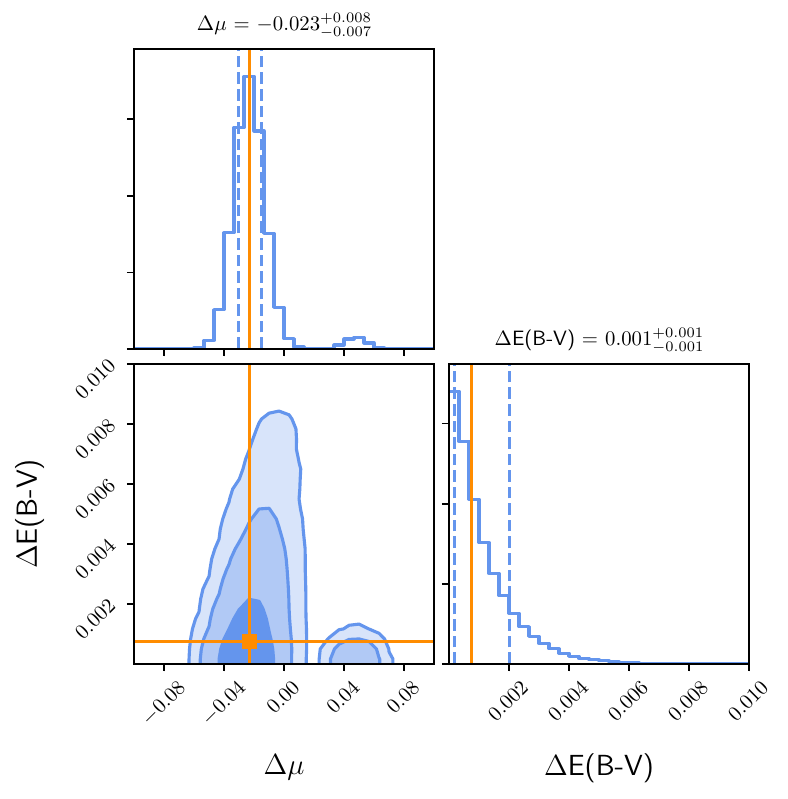}
  \caption{Left: Comparison of late-time spectra of SN 2013aa and SN 2017cbv at 361
    days past maximum light. Right: As in Figure 1 but for the 361 days phase.}
\end{figure}

\begin{figure}[H]
\includegraphics[width=0.45\textwidth]{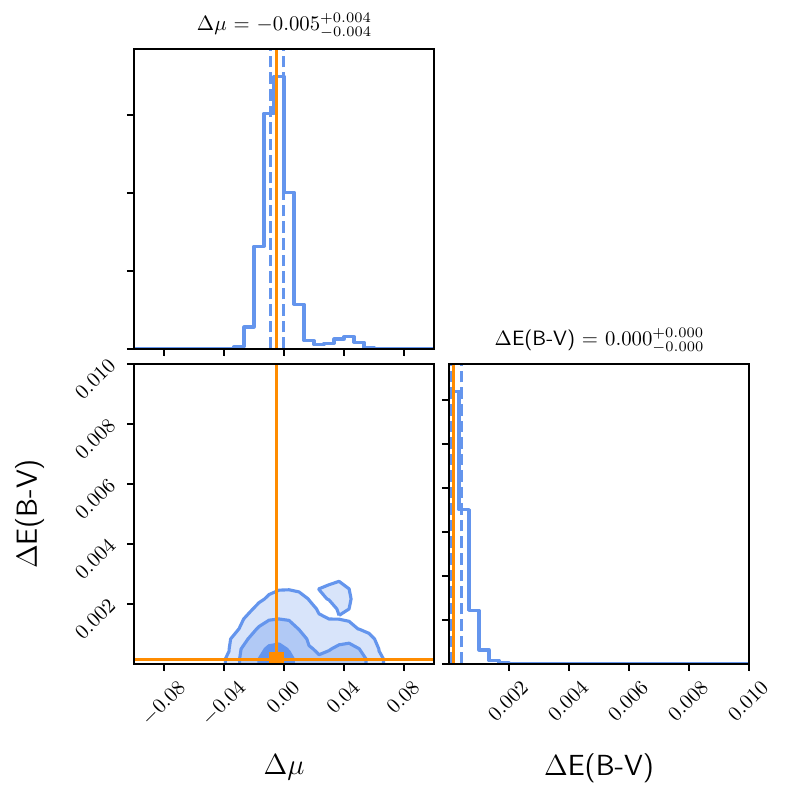}  
\includegraphics[width=0.45\textwidth]{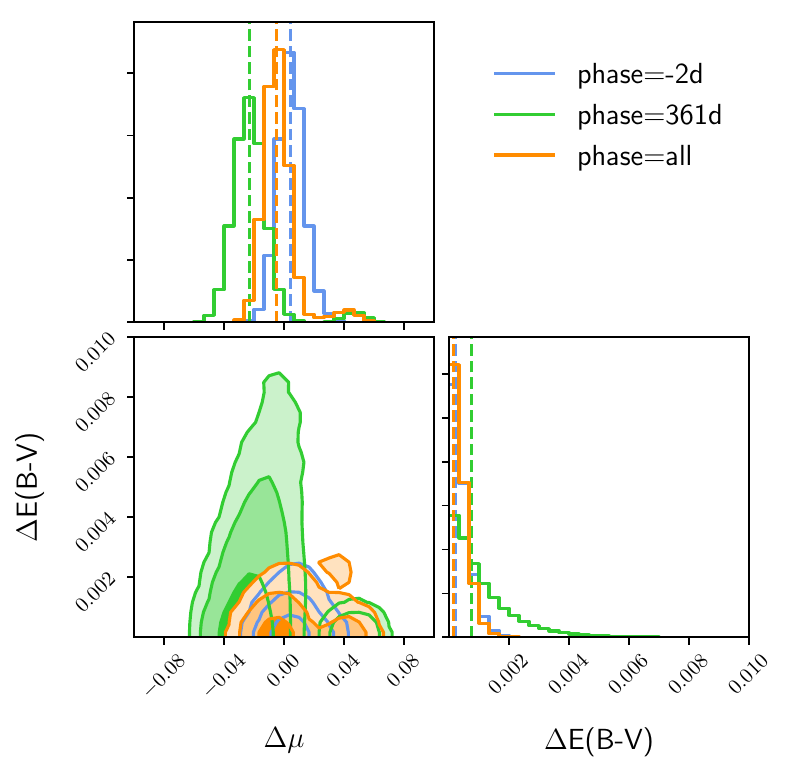}  

\caption{Left: Final joint result from
  early (-2d) and late time spectra (361d) of SN 2013aa and
  SN 2017cbv. Right: The results showing the corner plots with
  the 1$\sigma$, 2$\sigma$ and
  3$\sigma$ confidence regions favored by each phase and
  those from the joint computation.}
  \end{figure}

\bigskip

\noindent
 NGC 5643 will eventually become an important  step of reference
  in
 the distance ladder  for
supernovae belonging to the class of twins of
SN 2017cbv/SN 2013aa, hopefully reaching high
enough $z$, where peculiar velocities do no longer prevent to
reliably derive H$_{0}$. 

\bigskip

\subsection
    {Quantities that describe twin SNe Ia}

\bigskip

\noindent
Twin spectra of SNe Ia near maximum should have a similar
shape of the pseudo--continuum and similar pseudo--equivalent widths
(pWs) of the different lines. All that is required because otherwise
it is not possible to have a difference in magnitude of
$\sigma$ $=$ 0.04 in all filters.

\bigskip

\noindent
Other approaches that aim at addressing the similarity between spectra
use a quantification that mixes various lines to map 
the space of spectral diversity (Fakhouri et al. 2015; Boone  et al. 2016).
In our case, we want to give values of the most relevant lines that
characterize the physics of SNe Ia. Along this path, we
study the place where the different pWs of the lines fall 
in diagrams such as those
in Morrell et al. (2024).

\bigskip

\noindent
We have  measured  lines relevant 
for describing the similarity of the spectra.
The spectra of the
twins have very similar pseudo-equivalent widths (pWs) in the list
of lines called (see Morrell et all. 2024)
pW1 (Ca H \& K), pW2 (Si II $\lambda$ 4130 \AA \ ), pW3
(Mg $\lambda$ 4481; blended with Fe II),
pW4 (Fe II at $\sim$ 4600 \AA \, blended SII), pW5
(S II ''W'' $\sim$ 5400 \AA \ ), pW6 (Si II $\lambda$ 5972 \AA \ ),
pW7 (Si II $\lambda$ 6355 \AA \ ), pW8 (Ca II IR \AA \ ).
In particular, the line associated with Si II $\lambda$ 5972 \AA \
should be very similar in twin SNe Ia, with a discrepancy of less than
5$\%$. This line correlates with stretch ($\Delta m(B)_{15}$, s$_{BV}$), and
amongst SNe Ia of similar stretch, has to be almost the same for twins. 
In the pW6 over pW7 diagram the twins should fall in very close positions. 

\bigskip

\noindent
If the equivalent widths are similar, the ratios of the to Si II
at $\lambda$ 5972 \AA \ and $\lambda$ 6355 \AA \, $\Re_{Si II}$  should coincide.

\bigskip

\noindent
The shape of the pseudo--continuum should point to a similar
T$_{\rm eff}$ in order to overlap.

\bigskip

\noindent
In the nebular phase, the presence of stable Ni lines
should be similar as well as
the width and relevance of the lines.

\bigskip

\noindent
In Appendix A we present these checks on SN 2017cbv/SN 2013aa  and SN 2011fe. 

\bigskip

\section{The distance to NGC 7250}

\bigskip

\noindent
SN 2013dy in NGC 7250
has a decline rate similar to that of SN 2013aa and SN 2018cbv.
We have found enough similarities
to suggest that it belongs to the class of twins represented by SN 2017cbv
and SN 2013aa. It follows an evolution identical to those SNe Ia. The characteristics of this SN Ia can be found in Table 3. 

\bigskip

\noindent
SN 2013dy  was discovered a few hours after explosion.
This circumstance  made possible a very intense photometric and
spectroscopic follow up of this SN Ia (Pan et al. 2015).  This supernova
belongs to normal
SNe Ia which can be fitted with a W7 type model with solar metallicity
(Nomoto et al.  1984).

\bigskip

\noindent
We went further looking for differences and
similitudes with the two twin SNe Ia in NGC 5643. We were suprised to
see the similarities in spectral evolution. We have selected three
epochs to display this similitude: 2 days before maximum, 46 days
after maximum and 333 days after maximum.

\bigskip

\noindent
SN 2013dy is practically equal at late phases to SN 2017cbv (there are
no spectra of SN 2013aa at that phase to allow a comparison).
At very early phase
it is similar to both SN 2017cbv and SN 2013aa though slightly redder.
After maximum brightness is also identical to SN 2013aa.

\bigskip

\noindent
Pan et al. (2015) show a comparison of the UV flux of SN 2013dy
with that of SN 2011fe. The larger flux in SN 2013dy is consistent
with a larger amount of $^{56}$Ni than for SN 2011fe.
So it is its longer rise to B maximum,
which is 17.7 days in SN 2013dy against 17.6 days in SN 2011fe.

\bigskip

\noindent
SN 2013dy had significant reddening. The reddening in the Galaxy is
E(B-V)$_{MW}$ $=$0.14 mag (Schlafly \& Finkbeiner 2011), though there are
indications that the total reddening might be higher with a possible
lower R$_{V}$ than 3.1. We will address this once we made our comparison
of SN 2013dy with SN 2017cbv and SN 2013aa, those last ones  reddened with a
E(B-V)$_{MW}$$=$0.15 mag (Schlafly \& Finkbeiner 2011).

\bigskip
\bigskip

\begin{table}[ht!]
  \centering
  \caption{SN 2013dy}
  \begin{tabular}{lcc}
    \hline
    \hline
        {\bf SN 2013dy} &      &     \\
        RA, DEC$^{a}$ & 22:18:17.599 & +40:34:09.59 \\
        Discovery date$^{a}$ & $\dots$ & 2013-07-10 \\
        Phase (referred to maximum light)$^{b}$ & $\dots$ & -18 days \\
        Redshift$^{c}$ & $\dots$ & 0.0039 \\ 
        E(B-V)$_{MW}$$^{d}$ & $\dots$ & 0.15 mag \\
        $m_{B}^{max}$ $^{e}$ & $\dots$ & 13.229$\pm$0.010 mag \\
        $\Delta m(B)_{15}$ $^{e}$ & $\dots$ & 0.92$\pm$0.006 mag \\
        Phases of the spectra used & $\dots$ & -2, 46, 333 days \\
        \hline
  \end{tabular}
    \begin{tabular}{lll}
      $^{a}$Casper et al. (2013). & $^{b}$Zheng et al. (2013). &
      $^{c}$Schneider et al. (1992). \\
    $^{d}$Schlafly \& Finkbeiner (2011). & $^{e}$Pan et al. (2015).  &  \\  
    \end{tabular}
\end{table}

\begin{figure}[H]
  \centering
\includegraphics[width=0.53\textwidth]{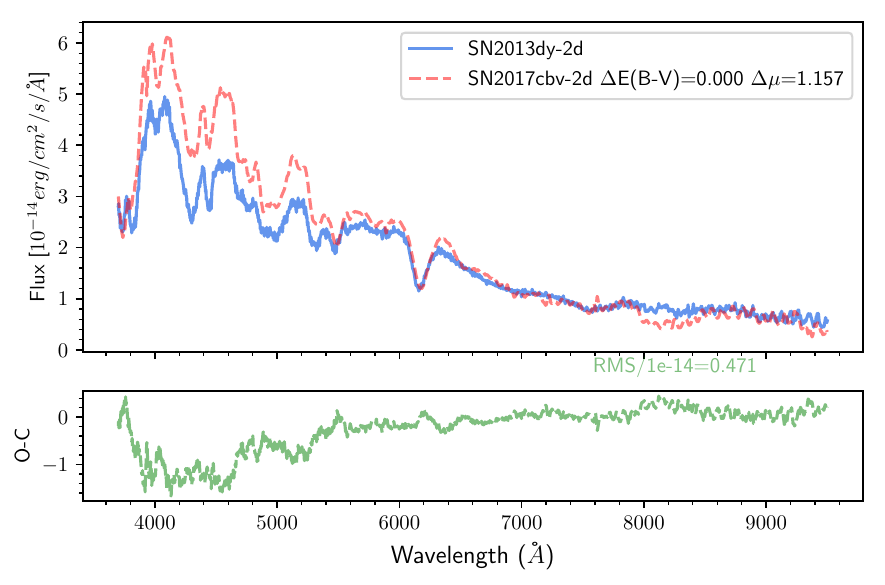}
\includegraphics[width=0.42\textwidth]{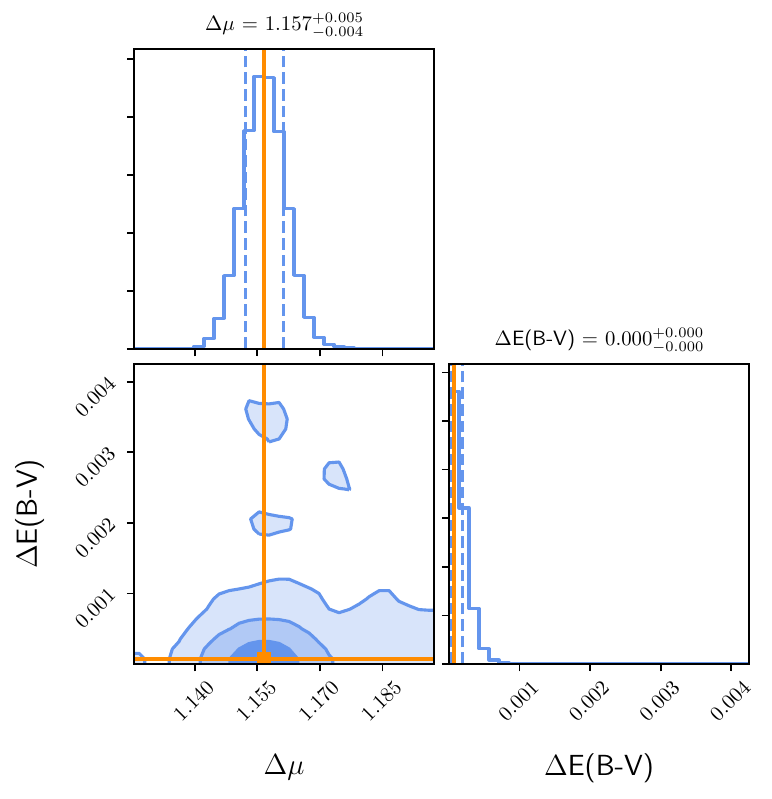}
\caption{Left: Comparison of early time spectra of SN 2013dy and SN 2017cbv  at
    -2 days before maximum light. Right: 1$\sigma$, 2$\sigma$, 3$\sigma$ contours for this epoch.}
  \end{figure}

\begin{figure}[H]
\includegraphics[width=0.53\textwidth]{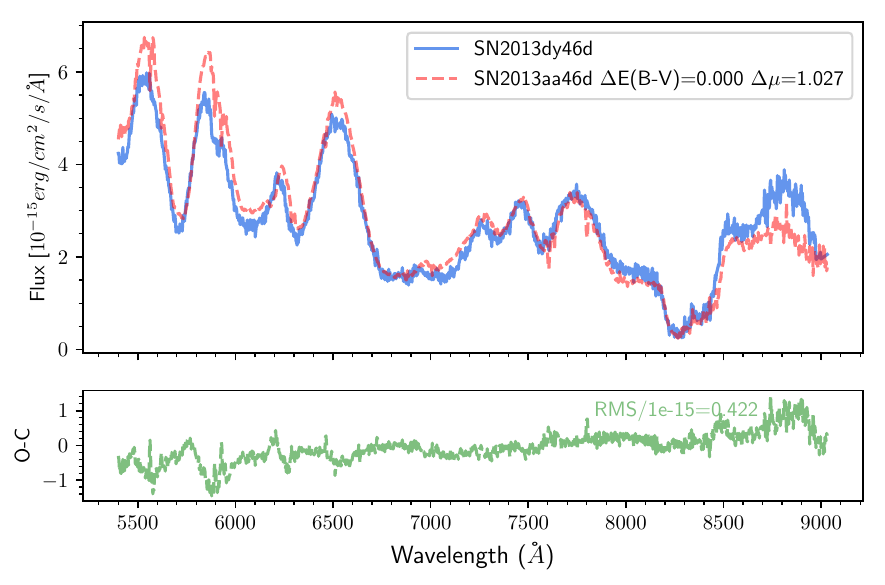}
\includegraphics[width=0.42\textwidth]{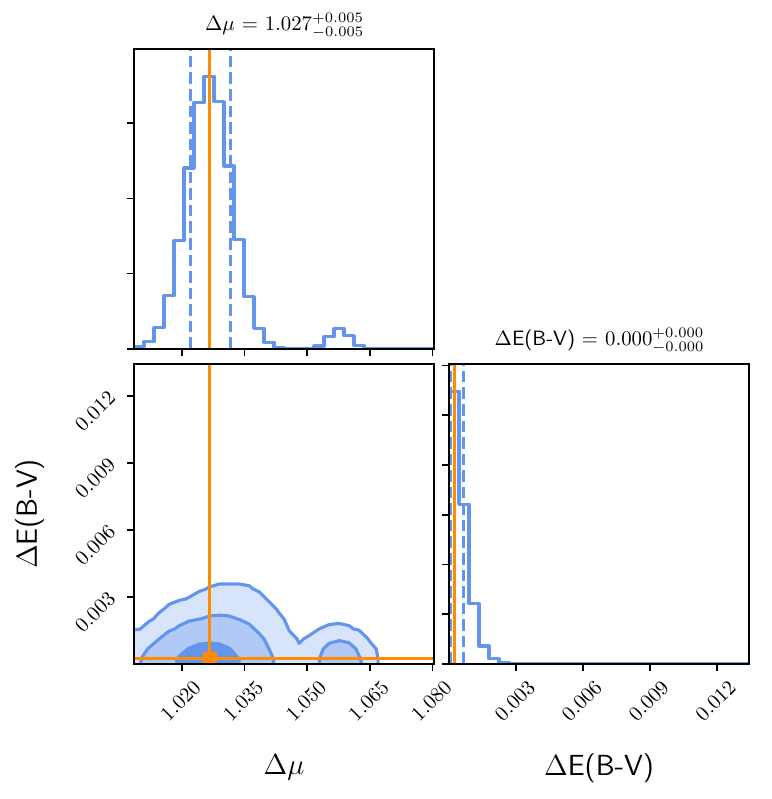}
  \caption{Left: Comparison of the spectra of SN 2013dy with SN 2013aa at 46 
    days past maximum light. Right: As above but for the 46 days phase.}
\end{figure}

\begin{figure}[H]
\includegraphics[width=0.53\textwidth]{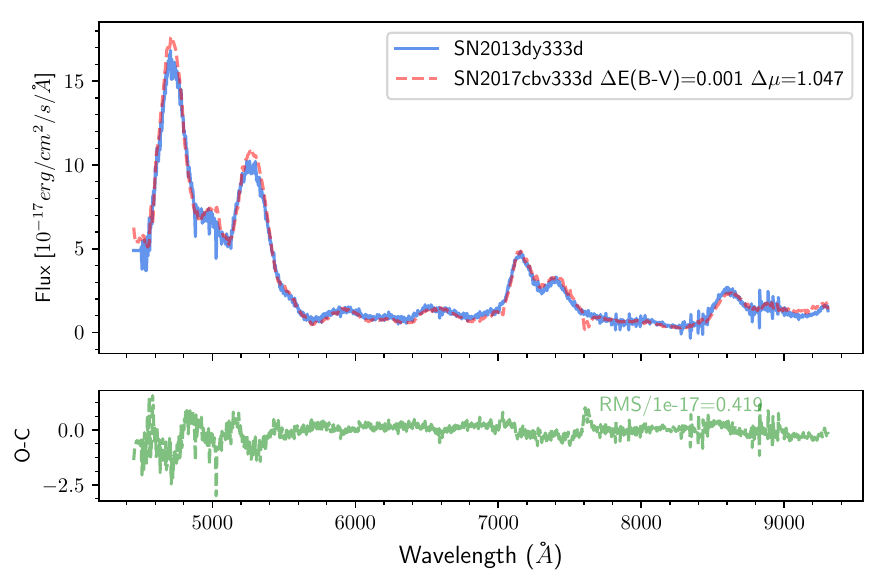}  
\includegraphics[width=0.42\textwidth]{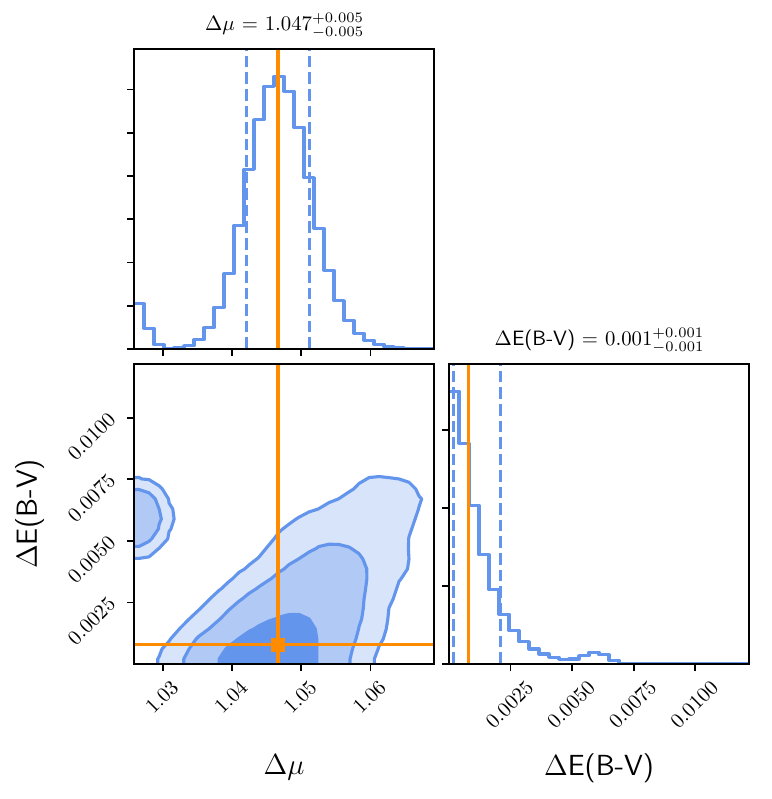}  
\caption{Left: Comparison of the spectrum of SN 2013dy and SN 2017cbv from
  late time spectra at 333 days. Right: Contours for the 333 days phase. }
  \end{figure}

\bigskip
\noindent
We see at early phase (2 days before maximum) (Figure 4) that SN 2013dy has
a less
steep spectrum (redder) than SN 2017cbv (and SN 2013aa). This is not due to
reddening (as can be seen in later phases) but to an intrisic slightly
different effective temperature of the spectrum. The spectral features are,
though, equal between SN 2013dy and its twins at early phases. One might
consider that this difference can be explained by an external factor,
as there is an early blue
excess both in spectra and the light curve in SN 2017cbv. 
It has been pointed out to the presence of a companion and to
an explosion from a double detonation initated at the surface of the star for
this supernova. In
the case of SN 2013aa, early data are missing, so it is not possible
to check whether there might have been indirect evidence for a companion.
But SN 2013aa is identical to SN 2017cbv, so it might have arisen in a
similar context.

\bigskip
\noindent
From our purpose of distance determination, we exclude the earliest
spectrum of SN 2013dy in our joint estimate of the distance of SN 2013dy, as
 there is a circumstancial factor only affecting the earliest epoch.
The spectrum of SN 2013dy achieves just a few days later an
amazing similarity to those of SN 2017cbv and SN 2013aa that last until
the nebular phase. We
obtain the derived distance to SN 2013dy from its comparison with phase
46 (Figure 5)  and 333 days past maximum (Figure 6). The combined posterior distributions can be seen in Figure 7. The results on the distance are
given in Table 4. 
One might note that $\Delta E(B-V)$ $=$ 0.0 $\pm $ 0.000.
The Pearson correlation coefficient r of the variables $\Delta \mu$ and
$\Delta E(B-V)$  for the 46 phase sample  $=$ -0.025, and for the 333 phase
$=$ 0.188, thus is very low. This might be explained by the fact that we
are moving around a $\Delta E(B-V)$ very low. 
SN 2013dy has
a similar total E(B-V) than its twins SN 2017cbv/SN 2013aa, and a reddening
of E(B-V) $=$ 0.01 for SN 2013dy would arise in the host galaxy
(E(B-V)$_{MW}$ is 0.14 for SN 2013dy).

\bigskip

\begin{figure}[H]
  \centering
\includegraphics[width=0.45\textwidth]{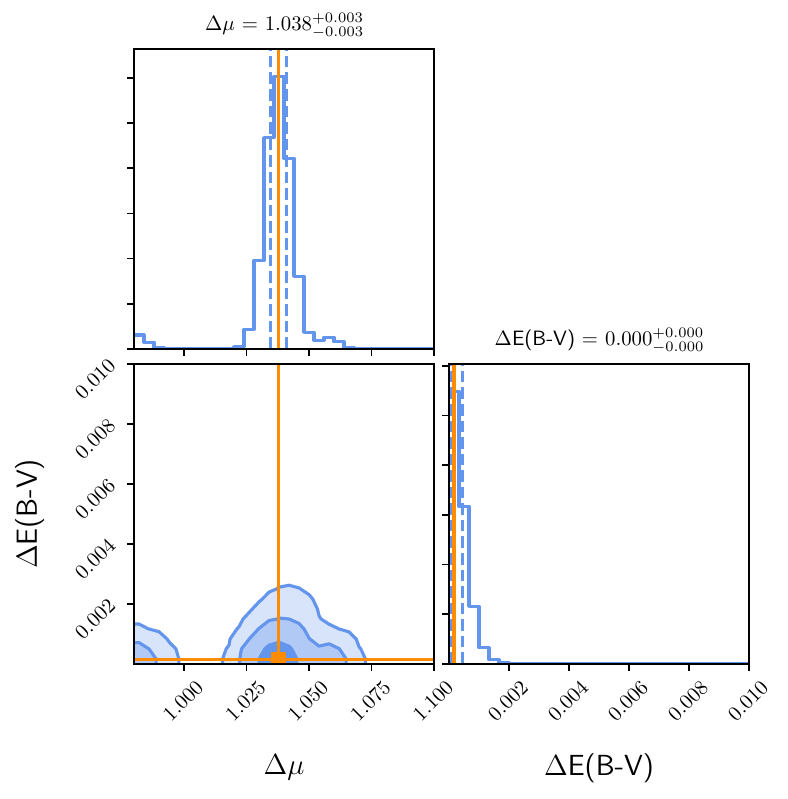}
\includegraphics[width=0.45\textwidth]{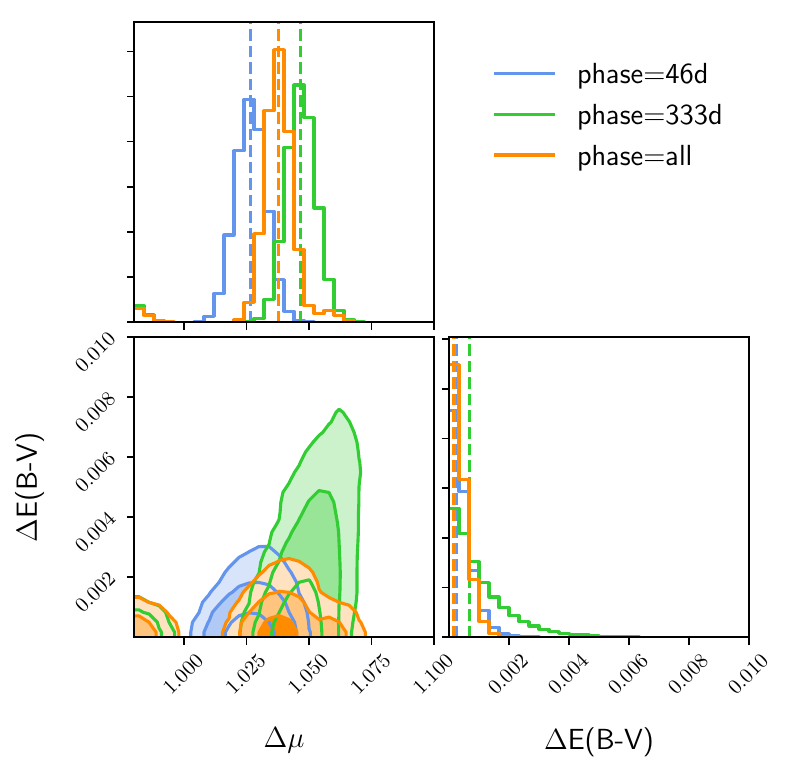}

\caption{As in Figure 3. Left: Final values favored by the joint phases both
   46 days after maximum and
   333 days after maximum of SN 2013dy with its twins. Right: Results for each
   phase and the joint phases.}
  \end{figure}

\bigskip

\begin{table*}[ht!]
  \centering
  \caption{The distance to NGC 7250}
  {\footnotesize
    \begin{tabular}{lccccc}
      \hline
      \hline
      SN & Host & $\mu$ & Error in $\mu$  & Source  \\
      \hline
      2013dy  & NGC 7250 & 31.518 & 0.003 + 0.1$^{*}$  & This work  \\
      2013dy  & NGC 7250 & 31.628 & 0.125 & Riess et al. (2022)
      (Cepheids) \\
      \hline
      {\bf  Previous step in the ladder} &    &   &   &      \\
      \hline
      2017cbv & NGC 5643 & 30.480 & 0.1 & Hoyt et al. (2021) \\
      2013aa  & NGC 5643 & 30.480 & 0.1 & Hoyt et al. (2021) \\
      \hline
      $^{*}$Error in the distance to NGC 5643 of 0.1 & & & & \\
      is twice the currently estimated error.  &  &  &  & \\
      See section about NGC 5643.  & &  &   & \\
\end{tabular}}
\end{table*}

\bigskip

\section{The distance to NGC 2525}

\bigskip

\noindent
In this Section we proceed to use twin SNe Ia of other type. In
particular, we are interested in the 2011fe--like group.  SN 2018gv
in NGC 2525 belongs to it. This allows us to determine the distance
to this galaxy, for which only old, rather unrealistic Tully--Fisher distance
measurements, plus a Cepheid measurement with substantial uncertainty are
currently available.

\bigskip

\bigskip

\noindent
We will now use the spectral twins method to estimate
the distance to NGC 2525 through the comparison between SN 2011fe and
SN 2018gv from early until nebular phases.  

\bigskip

\subsection{SN 2011fe in M101 as a reference}

\bigskip

\noindent
SN 2011fe was discovered, in the nearby spiral galaxy M101, in its very early 
phase and it was classified as a normal SN Ia (Nugent et al. 2011). It appears
to have been discovered within a few hours after the explosion and this helped
to put constraints on the progenitor system of the explosion (Nugent et al.
2011; Bloom et al. 2012). The pre-explosion Hubble Space Telescope image of
the SN 2011fe location ruled out luminous red giants and almost all helium
stars as the
mass-donor companion of the exploding WD (Li et al. 2011). In addition, the
early-time photometry of SN 2011fe (Bloom et al. 2012) set a limit to the
initial radius of the primary star, R$_{p} \leq 0.02 R_{\odot}$, as
well as a limit to the size of the companion star, R$_{c} \leq 0.1 R_{\odot}$.
Tucker \& Shappee (2023)  have reported imaging follow-up of SN 2011fe during
11.5 yrs after discovery, that setting strong constraints to a possible
single-degenerate progenitor systems. This
result appears consistent with the lack of interaction with a
non-degenerate companion in the Swift/UV light curves (Brown et al. 2012).
The limits on post-impact donors by Tucker \& Shappee (2023), together with
constraints from pre-explosion imaging, early-time radio and X-ray
observations, and nebular-phase
spectroscopy, essentially rule out all formation channels for SN 2011fe
invoking a non-degenerate donor star. This favours a double-degenerate
progenitor for SN 2011fe. 

\bigskip

\noindent
At maximum light, the color was 
(B$_{max}$ - V$_{max}$) = –0.07 $\pm$ 0.02 mag. The value
of (B$_{max}$ - V$_{ max}$) being –0.12 mag in normal SNe Ia, this might mean
that there is, in fact, some host galaxy extinction.
With the removal of the Galactic component, E(B--V)$_{MW}$ = 0.008 mag
(Schlafly \& Finkbeiner 2011), Pereira et al. (2013) estimated such reddening
as E(B - V)$_{host}$ = 0.026 $\pm$ 0.045 mag, by using the Lira relationship of
the colors in the postmaximum phase, in close agreement with the 0.032
$\pm$ 0.045 mag obtained by Zhang et al. (2016). 
Tammann \& Reindl (2011) obtained E(B--V)$_{host}$ = 0.030 $\pm$ 0.060 mag from
the B--V at the maximum of the light curves. The application of CMAGIC to 
SN 2011fe by Yang et al. (2020) gives a host galaxy extinction compatible with
zero, E(B - V)$_{host}$ = 0.0 $\pm$ 0.027 mag. Patat et al. (2013)
studied, from multi–epoch high resolution spectroscopy of SN 2011fe, the
reddening based on
the absorption systems of Ca II and Na I towards the supernova. They inferred 
a host galaxy reddening from the equivalent
width (EW) of  Na I of E(B-V) $=$ 0.014 mag and they derive a
Galactic reddening of E(B-V)$_{MW}$ $=$ 0.01 mag  (similar to tha from
Schlafly \& Finkbeiner 2011). So, there is a total E(B-V)$=$ 0.024 mag from
this last reference. We adopt such value as a most accurate guess, though the
absolute value is not something needed for the application of the method. The
reddening relative to its twin will be determined by the MCMC. 

\bigskip

\noindent
Concerning the luminosity decline parameter of SN 2011fe,
Burns  (2023, private communication) measured a
$\Delta m_{15} (B)$ = 1.07$\pm$0.006 mag. The basic data for SN 2011fe are
given in Table 5. 

\bigskip

\noindent
The derived Fe abundance in the outermost layers of the ejecta is consistent
with the
metallicity at the SN site in M101 ($\sim$ 0.5 Z$_{\odot}$: 
Mazzali et al. 2015). This produces differences in the spectrum as
compared with the W7 model, which has solar metallicity, with higher
UV flux and affecting as well the blue spectrum at early phases. At
post--maximum phases, save
differences in the UV part of the spectrum, the rest of the wavelengths
converge to a model based on solar metallicity such as W7. 
This makes twin pairing with the SN 2011fe-like SNe to be better in the
optical and improving after maximum, when the photosphere has receded
to the SN core.
  Several modeling approaches give, for SN 2011fe, an amount of $^{56}$Ni of
  0.5 M$_{\odot}$ synthesized
  in the explosion. Pereira et al. (2013), from the bolometric light curve
  of the supernova obtain 0.53 $\pm$ 0.11 M$_{\odot}$ of $^{56}$Ni.
  Zhang et al (2016), using as well the bolometric light curve, estimate
  around $\sim$ 0.57 M$_{\odot}$ of $^{56}$Ni. Mazzali et al. (2015)
  derived a mass of $^{56}$Ni $\sim$ 0.47 $\pm$ 0.05 M$_{\odot}$ and
  a stable iron mass of $\sim$ 0.23 $\pm$ 0.03 M$_{\odot}$ for SN 2011fe,
  based on modeling of the nebular spectra.
  A centrally ignited SN Ia in a
  Chandrasekhar-mass model similar to W7 (with production of a slghtly lower
  mass of
  $^{56}$Ni) seems to agree with the chemistry of this SN.

\bigskip
  
\noindent  
  Since SN 2011fe turns to be a kind of ``template'' for a class of normal
  SNe Ia, this nearby supernova teaches us the level of
  $^{56}$Ni for that class.

\bigskip

  \noindent
  A distance to M101 (the host galaxy of SN 2011fe) as that provided by the
  TRGB method is also in good agreement with the observed flux of the SN. 
Here, however, we basically deal with the difference of distances between
SN 2011fe and SN 2018gv, although SN 2011fe being the anchor (the previous step
of reference in the distance ladder) here, its
distance enters in the final result.

\bigskip

\begin{table}
  \centering
  \caption{SN 2011fe and SN 2018gv}
  \begin{tabular}{lcc}
    \hline
    \hline
        {\bf SN 2011fe} &      &     \\
        RA, DEC$^{a}$ & 14:03:05.810 & +54:16:25.39 \\
        Discovery date$^{a}$ & $\dots$ & 2011-08-24 \\
        Phase (referred to maximum light)$^{b}$ & $\dots$ & -20 days \\
        Redshift$^{c}$ & $\dots$ & 0.0012 \\ 
        E(B-V)$_{MW}$$^{d}$ & $\dots$ & 0.008 mag \\
        $m_{B}^{max}$ $^{e}$ & $\dots$ & 9.983$\pm$0.015 mag \\
        $\Delta m(B)_{15}$ $^{e}$ & $\dots$ & 1.07$\pm$0.06 \\
        Stretch factor $s_{BV}^{D}$$^{b}$ & $\dots$
            & 0.919$\pm$0.004 \\
        Phases of the spectra used & $\dots$ & 9, 289, 344 days \\
        \hline
            {\bf SN 2018gv} &     &    \\
            RA, DEC$^{f}$ & 08:05:34.580 & -11:26:16.87 \\
            Discovery date$^{g}$ & $\dots$ & 2018-01-15 \\
            Phase (referred to maximum light)$^{f}$ & $\dots$ & -16 days \\
            Redshift$^{f}$ & $\dots$ & 0.0053 \\
            E(B-V)$_{MW}$$^{d}$ & $\dots$ & 0.051 mag \\
            $m_{B}^{max}$ $^{e}$ & $\dots$ & 12.8$\pm$0.015 mag \\
            $\Delta m(B)_{15}$$^{f}$ & $\dots$& 0.983$\pm$ 0.15 \\
            Stretch factor $s_{BV}^{D}$$^{b}$ & $\dots$
   & 0.937$\pm$0.031 \\
            Phases of the spectra used & $\dots$ & 9, 289, 344 days \\
            \hline
  \end{tabular}
    \begin{tabular}{lll}
    $^{a}$Waagen (2011). &  $^{b}$Richmond \& Smith (2012) & $^{c}$Cenko et al.
    (2011) \\
    $^{d}$Schlafly \& Finkbeiner (2011).&  $^{e}$Burns
    (private communication, (2023). & $^{f}$Yang et al. 2020. \\
    $^{g}$Itagaki (2018). &      &      \\
    \end{tabular}
\end{table}

\subsection{2018gv}

\bigskip

\noindent
SN 2018 gv is a SN 2011fe-like supernova. The explosion was fairly symmetric
and the
spectra were similar all along the different phases. The supernova was
discovered on 2018-01-15 by K. Itagaki in the outskirts of the host galaxy
NGC 2525 (Itagaki 2018), a barred spiral galaxy at $z$ = 0.00527 (de
Vaucouleurs et al. 1991).

\bigskip
    
\noindent
    The Galactic reddening in the direction of SN 2018gv is E(B-V)$_{MW}$ =
    0.051 mag, according to the extinction map by Schlafly \& Finkbeiner (2011).
    It is appreciably higher than that of SN 2011fe. The host galaxy reddening
    should be very low, given the position of SN 2018gv in the outskirts of its
    host galaxy. The test for reddening from the Lira relation
    gives that same amount, for the total reddening, as for just the
    Galactic one. The application of the
    CMAGIC method in Yang et al. (2020) gives a host reddening of
    E(B-V)$_{host}$ = 0.028 $\pm$ 0.027 mag, thus being compatible with zero.

    \noindent
    The basic information on SN 2018gv can be
found in Table 5.
The early-time spectra of SN 2018gv show strong similarity in most respects
with those of the normal
Type Ia SN 2011fe (Yang et al. 2020).
These authors demonstrate
that SN 2018gv resembles SN 2011fe for the first 100 days and exhibits a low
Si II velocity gradient in the days after peak brightness. The
observation of
low continuum polarization overlaid by significant line polarization would be
inconsistent with an asymmetric explosion. They suggest an amount
of $^{56}$Ni of 0.56 $\pm$ 0.08 $M_{\odot}$  from the bolometric light
curve (Yang et al. 2020). Graham et al. (2022) show the even stronger 
late-time similarity
between SN 2011fe and SN 2018gv. They fully deserve to be named 
``nebular twins': at the same phases in the nebular evolution they have 
extremely similar spectra. For the application of
our method, the likeness of the spectra of twin SNe Ia can not just
consist of the fact that two nebular spectra might be, at some particular close
times, similar. It should happen at exactly
the same nebular stages. Being similar through the whole nebular phase 
indicates closely resembling Ni cores, which are the sources of the
luminosity of the supernovae at those epochs. 

\bigskip 

\noindent
As stated above, the two SNe Ia are very similar at early times, and that
mostly persists until the nebular phase, with a few noticeable
differences, however. So, for instance, the line
width measurements suggest a slight difference in central density,
though both densities are high enough to form stable Fe-group elements in the
innermost regions of the core.

\begin{figure}[H]
\includegraphics[width=0.53\textwidth]{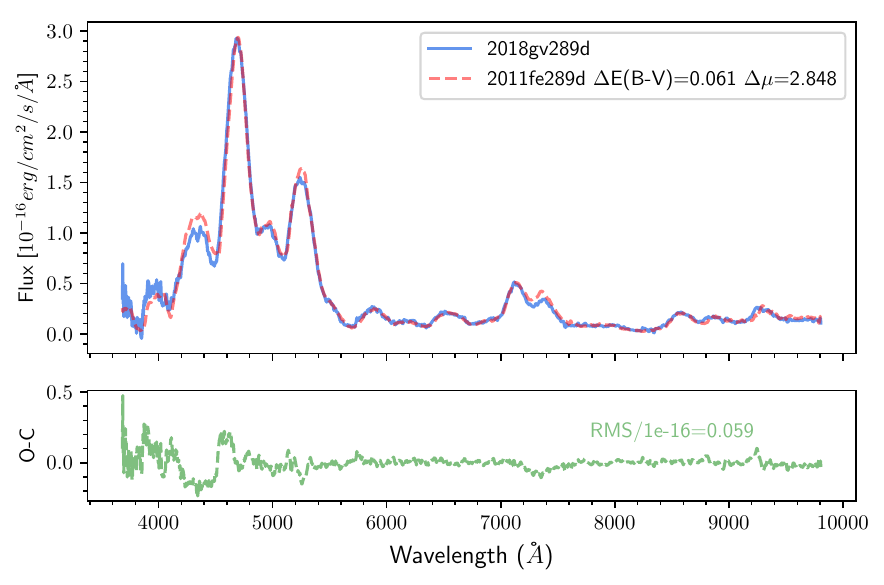}
\includegraphics[width=0.42\textwidth]{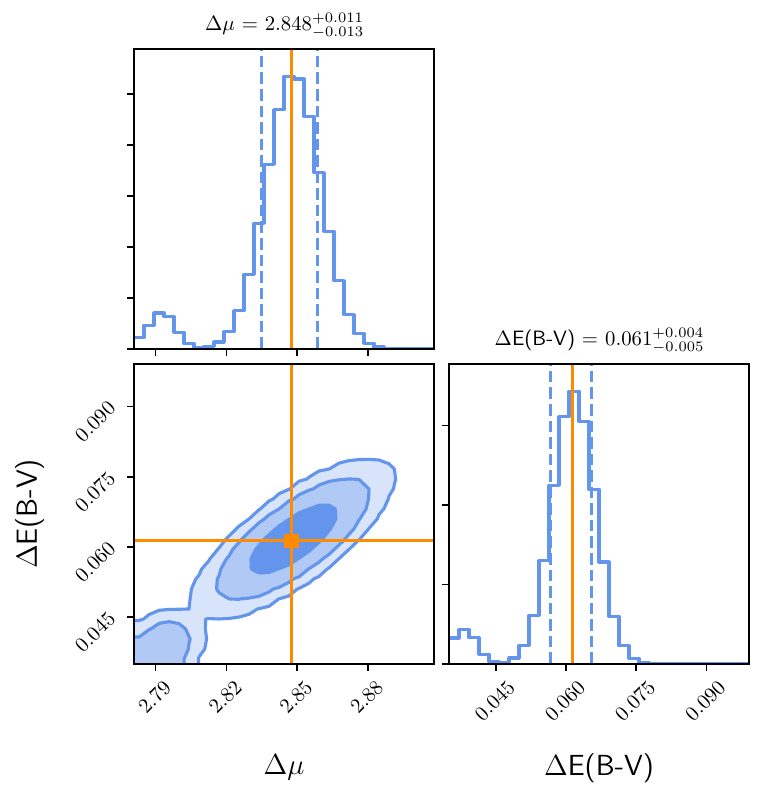}
  \caption{Left: Spectra of the
    twins SN 2011fe and SN 2018gv at exactly the same phase of 289 days past
    maximum light. Right: Corner plot as in previous figures but for SN 2018gv
    at 289 days.}
\end{figure}

\begin{figure}[H]
\includegraphics[width=0.53\textwidth]{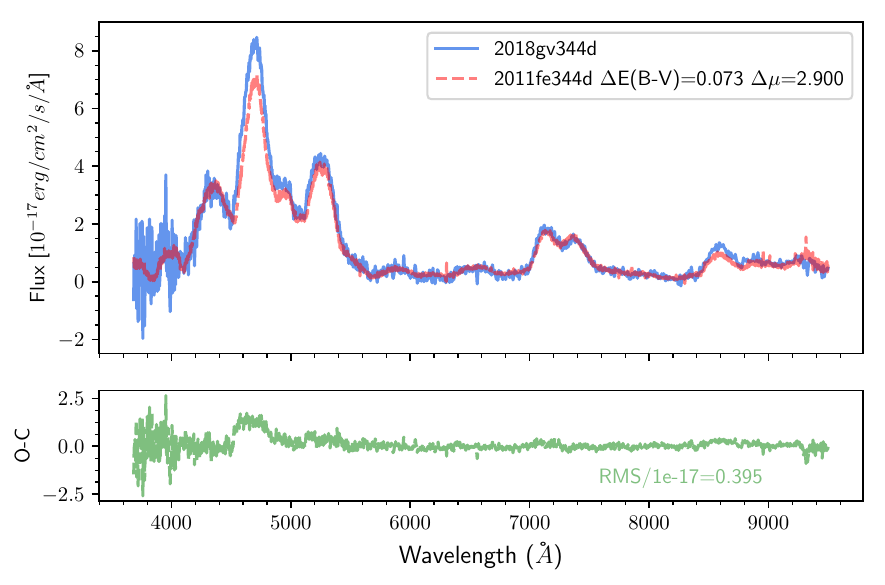}
\includegraphics[width=0.42\textwidth]{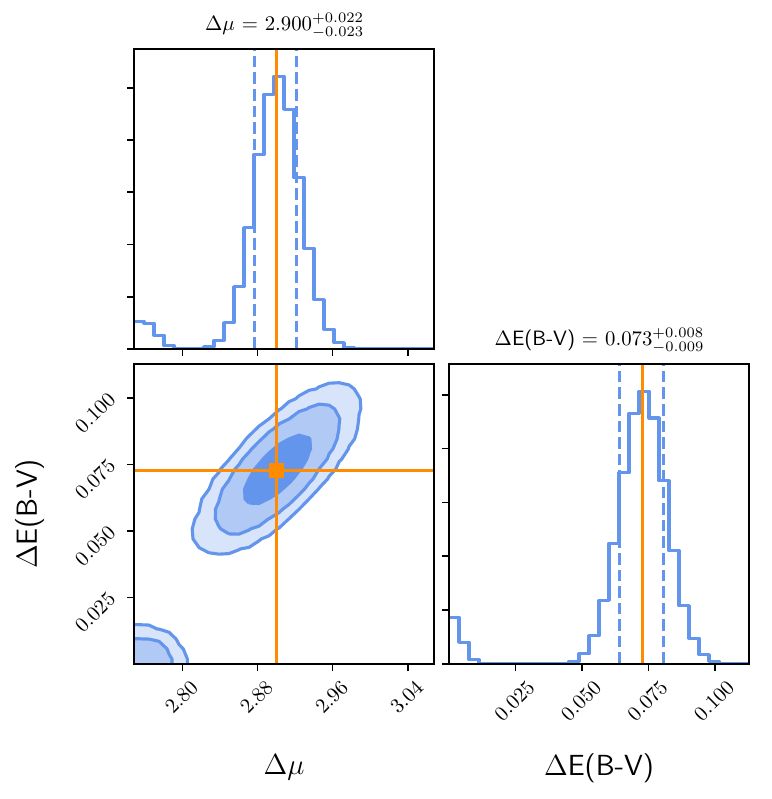}
  \caption{Left: Spectra of the
    twins SN 2011fe and SN 2018gv at exactly the same phase of 344 days past
  maximum light. Right: As above but for 344 days past maximum.} 
\end{figure}

\begin{figure}[H]
  \centering
\includegraphics[width=0.45\textwidth]{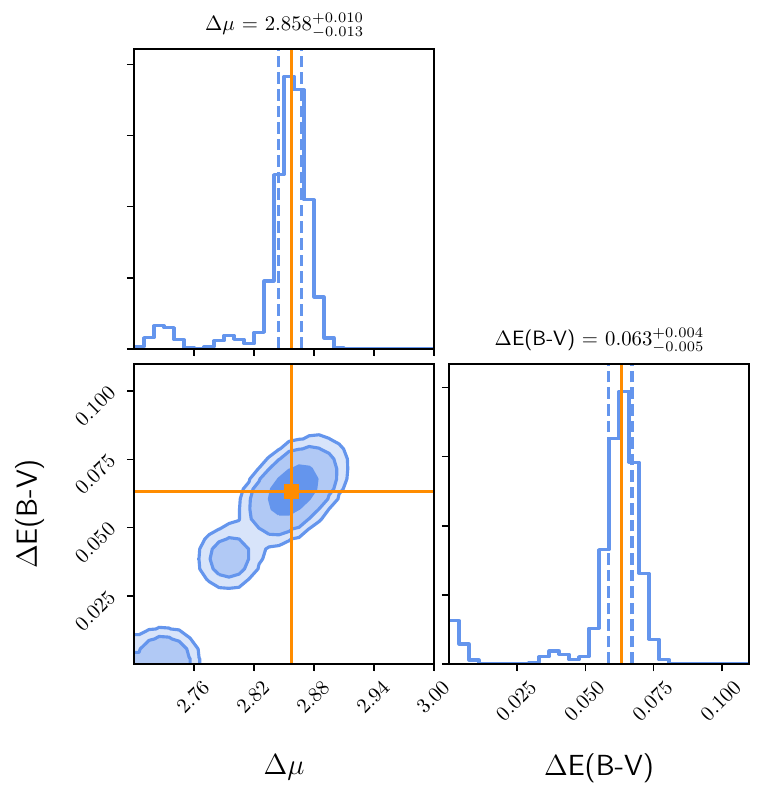}
\includegraphics[width=0.45\textwidth]{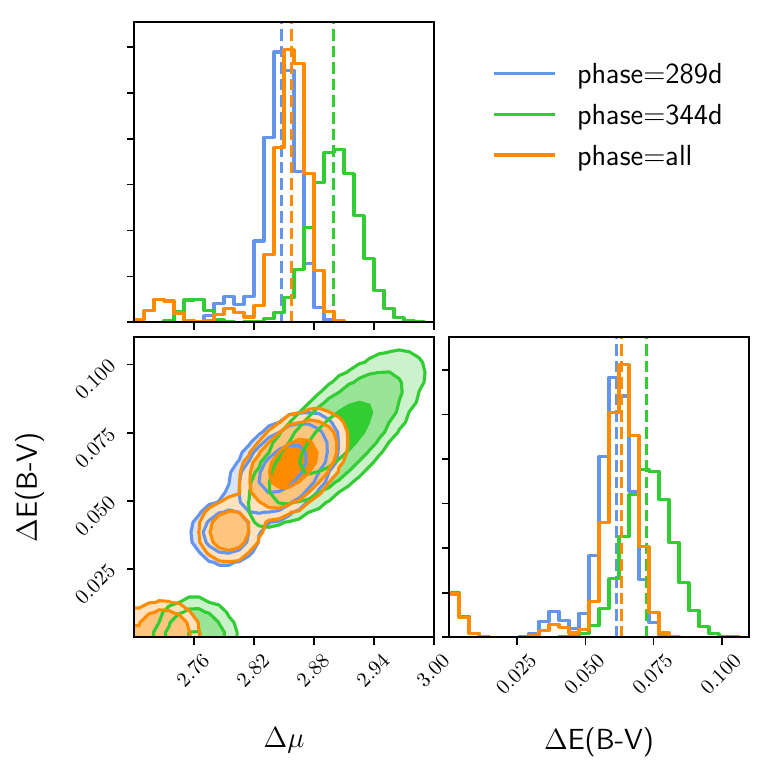}
\caption{(As in Figure 3 and 7). Left: Result from the joint comparison of the two phases. Right:
  Individual phases  and joint phases confidence regions
  from the comparison at 289 days after maximum and 344 days after maximum of SN 2018gv and SN 2011fe.}
\end{figure}

\subsection{$\Delta E(B-V)$ and distance}

\bigskip

\noindent
SN 2011fe and SN 2018gv were not heavily obscured by dust, as they occurred in
the outskirts of their host galaxies. The Galactic redenning towards their
direction has been measured, as stated above,
by Schlafly \& Finkbeiner (2011) for both SNe Ia.

    \bigskip

    \noindent
    To compare the spectra of the two SNe Ia, we have dereddened
    and deredshifted them both.
    SN 2011fe has been dereddened by 0.024 and SN 2018gv by
    0.05 mag. The quantity $\Delta$ E(B - V) is the amount of reddening that
    we would need to add to SN 2018gv to get the best fit between the
    spectra of the two SNe. The result, $\Delta$ E(B - V), indicates whether
    E(B–V) might
    be higher in SN2018gv than in SN2011fe.

\bigskip

\noindent
The results concerning the distance are, as already stressed, a relative
distance
    between SN 2018gv and SN 2011fe. SN 2011fe, being one of the nearest
    and best observed SNe Ia and a prototype ``normal'' SN Ia, is our
    first SN Ia that can work as an anchor for the extragalactic distance
    scale. 

    \bigskip

    \noindent
    We search for the best match between SN2018gv and SN2011fe using only two
    free parameters, the reddening value difference, $\Delta$ E(B-V), and
    modulus difference $\Delta \mu$. We have used the similarity
    between the spectra of SN2011fe and SN2018gv at three different phases
    and we perform a fit using the likelihood function defined in [1] (Section
    3).

    \bigskip

    \noindent
  We run again
  the Markov Chain Monte Carlo with 32 walkers and 10,000 steps each,
  implemented in
  EMCEE (Foreman-Mackey et al.
  2013), sufficient to get a statistically significant result. In each MCMC
  sample our code is able to deredden the reference spectrum
  (that of SN 2011fe) with a given value of E(B-V) and size the parameters
  $(\Delta E(B-V)$ and
  $\Delta \mu$),
 relative to SN 2018gv. We adopted uniform priors on the $\Delta D$
($\mathcal{U}[15, 21]$ Mpc)
 and reddening $\Delta E(B - V) (\mathcal{U}[0.0,0.8]$ mag). In Figure 8 and
 Figure 9, we
 show the posterior distributions for two different epochs and in Figure 10
 the one corresponding to the combined regions. At the end,
we found that the
early spectrum we were going to use
was not reliable as there were different reductions for
the same epoch with incompatible spectral shape 
in the {\it WISeREP} (see the above database to check this point).
We should note that here the Pearson correlation
coefficient r is significantly higher than for SN 2013dy, manifesting
the obvious implications that a change of $\Delta (E-V)$ has in $\Delta \mu$
in this nearly identical spectra. 
As we see, the results
    favour  $\Delta E(B - V)$ = 0.063 $^{+0.004}_{-0.005}$ mag more in SN
    2018gv than in SN 2011fe, and
    $\Delta \mu$ = 2.858 $\pm$ 0.013 mag. 
The distance factor between SN 2018gv and SN 2011fe is 3.71: $D_{2018gv} =
3.71 \times D_{2011fe}$. Those are sumarized in Table 6 and 8.

\subsection{ M101}

\bigskip

\noindent
     The distance to M101 has been on debate for decades.
      Several methods  have been used to estimate the distance to M101.
      But finally
      there seems to be an agreeement within a short range of discrepancy
      by two methods: the TRGB method and the Miras variable Period Luminosity
      relation. In addition, those methods are in agreement with what has been 
      learnt about SN 2011fe in this galaxy.

\bigskip
      
\noindent
At this moment, the estimates using Cepheids ({\it SH0ES}) differ
from that obtained through TRGB and the Miras variables.
But the value provided by Cepheids to the distance to this galaxy has
been changing along the years.
Riess et al. (2016) found a distance modulus of 29.135$\pm$0.045 mag
and, in Riess et al. (2022), a reanalysis of the data gave
29.194 $\pm$ 0.039 mag, which place M101 at practically 7 Mpc.

\bigskip

\noindent
Tammann
\& Reindl (2011) pointed out that the problem of the Cepheids in M101 is that
those in the  outer,
metal-poor field (Kelson et al. 1996) and those in two inner, metal-rich fields
(Shappee  \& Stanek 2011), yield discordant distances.

    \bigskip

\noindent
 The TRGB distance to M101 value in Freedman et al (2019)
  is coming from Beaton et al.
  (2019).  This  distance modulus is  $\mu$ = 29.07 $\pm$
  0.04 (stat) $\pm$ 0.05 (sys) mag, which corresponds to a physical distance
  $D$ = 6.51 $\pm$ 0.12 (stat) $\pm$ 0.15 (sys) Mpc. 
  Huang et al. (2024) recently
  obtained a distance measurement to M101 using Miras variables.
  The value by this method
is $\mu_{M101}$ $=$ 29.10 $\pm$ 0.06 mag in  very close agreement
with the TRBG distance to M101  and consistent
with our expected luminosity rank  of SN 2011fe (Appendix B).
This is an important step for the improvement of the cosmic 
distance ladder.

\bigskip

\noindent
     The Mira variables
       value places M101 at 6.60$\pm$ 0.19 Mpc, the TRGB
       at 6.51 $\pm$ 0.27 Mpc very much in agreement
      with the SN 2011fe at 6.5 $\pm$ 0.15 Mpc
      (Appendix A).

      \bigskip

\noindent
    We can say that M101 is  nowadays a firm step in the cosmic ladder.
    It consolidates its distance in the                                         range of  6.4$ <$ d $<$ 6.7 Mpc centered at 6.53 Mpc with a mean value for
    $\mu$ $=$ 29.075 $\pm$ 0.068 mag. The value encompasses the
    most recent coincident results by various methods: those mentioned above
    and summarized in Table 6. 
    This convergence  not only allows the calibration  of many other
    distances for SNe Ia
    of the twin type that happened in M101 but
    also cross-checks with SNe Ia host galaxies of other twin types.

\bigskip

\begin{table}[ht!]
  \centering
    \caption{Distance moduli for SN 2018gv and SN 2011fe}
  \begin{tabular}{llc}
    \hline
    \hline
    2018gv & $\mu$ = 32.067$\pm$0.10 & Cepheids$^{1}$ \\
    2018gv & $\mu$ = 31.933$\pm$0.069 & This work \\
    \hline
    2011fe & $\mu$ = 29.135$\pm$ 0.045& Cepheids$^{2}$ \\
    2011fe & $\mu$ = 29.194$\pm$0.039 & Cepheids$^{1}$ \\
    2011fe & $\mu$ = 29.07$\pm$0.09 & Tip of the Red Giant Branch$^{3}$ \\
    2011fe & $\mu$ = 29.10$\pm$0.06 &  Mira variables$^{4}$ \\
    2011fe & $\mu$ = 29.06$\pm$0.05 &  Appendix A \\
    2011fe & $\mu$ = 29.075 $\pm$ 0.068 &  Obtained M101 ladder step \\
\hline
  \end{tabular}
  \begin{tabular}{lll}
   \ $^{1}$Riess et al. (2022). & $^{2}$ Riess et al. (2016). &\\
   \ $^{3}$Freedman et al. (2019). & $^{4}$ Huang et al. (2024). &\\
    \end{tabular}
\end{table}

\bigskip

\noindent
We can use the convergent  distance value to
M101  $\mu$ $=$ 29.075 $\pm$ 0.068 mag in our SNe Ia twins
distance ladder. 
$\Delta \mu$ between SN 2011fe and SN 2018gv
is 2.858$\pm$ 0.013  mag in our results (see Figure 10).
So, we derive $\mu$ = 31.933 $\pm$ 0.069 mag for NGC 2525.

\bigskip

\noindent
The determination by {\it SH0ES} of the distance to NGC 2525 corresponds to  a
distance
modulus of $\mu$ = 32.067 $\pm$ 0.1 mag, which is a distance of
25.90$\pm$1 Mpc (Riess et al. 2022). 
The distance factor between SN 2018gv and SN 2011fe becomes
3.75 by Cepheids. Though it is likely that the Cepheids
  distance to NGC 2525 will be revised
with new data from {\it JWST}, as has been done for other galaxies.

\bigskip

\noindent
In our determination of the distance to NGC 2525, the distance modulus obtained
corresponds to
a distance of 24.35$\pm$0.77 Mpc when adopting the distance modulus
for
M101  (see Table 6). The distance factor of SN 2018gv
in relation to SN 2011fe is 3.71, then. There is agreement within 1$\sigma$
with the {\it SH0ES} value 
because the error in NGC  2525 with Cepheids is quite large\footnote{
  The TRGB distance modulus to NGC 2525 by the {\it SH0ES}  collaboration 
    has been revised
    just very recently and it is now 31.81 $\pm$ 0.08 mag. It is compatible with
    our measurement, and shorter than in Riess et al. (2022), as
    we expected.}.

  \subsection{NGC 5643}

\bigskip

\noindent
It might have seemed in Section 3 and Table 4 that we made an arbitrary choice
in refering the distance of NGC 7250  to the NGC 5643 by Hoyt et al. (2021),
and that we could have done other choice like the value provided by Cepheids in
Riess et al. (2022). However, the distance values published
in Riess et al. (2022) 
seem to have been abandoned in the case of NGC 5643 and other galaxies.
There are
several papers where a new Cepheid distance is given, very close to the
distance by Hoyt et al. (2021), and a previous distance is quoted with the method
of Cepheids, but it is not the distance in Riess et al. (2022).
It is a distance very close to the TRGB value by Hoyt et al. (2021).
Table 3 in Riess et al.
(2024a) presents a {\it JWST+ HST}
value of the modulus to NGC 5643 of $\mu$ $=$ 30.52
$\pm$ 0.02 mag
and a comparison value with HST ({\it SH0ES})  of $\mu$ $=
$ 30.518
$\pm$ 0.033 mag. The value $\mu$ $=$ 30.57
$\pm$ 0.05 mag from Riess et al. (2022)
was the reference value of their major paper with 37 host galaxies of
SNe Ia in 2022. In
Riess et al. (2024a) there is no mention to this value, while it is argued
that ''JWST observations reject unrecognized crowding of Cepheid photometry
as an explanation for the Hubble Tension at 8$\sigma$ confidence. ''

\bigskip

\noindent
A compilation of the values to NGC 5643 with the Cepheids  and the TRGB
methods is given in Table 7. 

\bigskip

\begin{table}[ht!]
  \centering
      \caption{The distance to NGC 5643}
          \begin{tabular}{lccc}
      \hline
      \hline
      Galaxy & Distance modulus (mag) & Method & Reference \\
      \hline
      NGC 5643 &  30.480$\pm$0.1 &  TRGB  & Hoyt et al. (2021) \\
      NGC 5643 &  30.42$\pm$0.07 &  TRGB     & Anand et al. (2024) \\
      NGC 5643 &                 &           & Anand et al. (2024) \\
      \hline
      NGC 5643 &  30.570$\pm$0.050 &  Cepheids & Riess et al. (2022) \\
      NGC 5643 &  30.52$\pm$0.02 &  Cepheids (JWST+HST) & Riess et al. (2024a) \\
      NGC 5643 &  30.518$\pm$0.033 & Cepheids (HST) & Riess et al. (2024a) \\
      \hline
      NGC 5643 &  30.48$\pm$ 0.065 & SN 2013aa/SN 2011fe &    \\
      \hline
      NGC 5643 &  30.48$\pm$0.065 & & Obtained NGC 5643 ladder step   \\
     
      \hline

    \end{tabular}
    \end{table}

\bigskip

\noindent
There is another paper using TRGB by Anand et al. (2024)
that suggests a new distance modulus to NGC 5643 very similar to that by
Hoyt et al. (2021). Such distance is in fact $\mu$ $=$ 30.42 $\pm$ 0.07 mag
and different from $\mu$ $=$ 30.57 $\pm$ 0.05 mag from  Cepheids in 
Riess et al (2022) ({\it SH0ES} value). 
The Anand  et al. (2024) distance modulus is in fact
the one found by Anand et al. (2022)
reanalysing the data taken by Hoyt et al. (2021) for NGC 5643. 
Anand et al (2024)  in their Figure 11 shows the
$\mu$ value obtained by Hoyt et al. (2021) and the one from their
reanalysis in Anand et al. (2022), and they
point that $< TRGB -Ceph >$ is very small.  However, that would not be
the case if the data would have been taken from Riess et al. (2022). 

\bigskip

\noindent
We have examined the
distance to SN 2013aa/SN 2017cbv based in their relation towards SN 2011fe.
We find that the value provided by Hoyt et al. (2021)  coincided with that
from the relation betweeen SN 2013aa/SN 2017cbv SNe Ia and SN 2011fe types.
There is a consistent ladder step
reflecting the amount of radiactive material sythesized in the explosion of
this type versus other ones like SN 2011fe
(see next section).  Given the coincidence in distances as shown in Table 7 with
the TRGB and within the recent Cepheid values, we think that there is
a good based ladder step in NGC 5643 at a $\mu$  of 30.48$\pm$0.065 mag.

\bigskip

\noindent
It is good to find convergence in some new
distance values for a few galaxies between
Cepheids and TRGBs, however the previous larger data set with important
disagreements should be mentioned as well.
Otherwise, there is a  loss of  credibility in the conclusions.

\bigskip

\noindent
We note, in a previous version of the paper, that
the Cepheid distance modulus for NGC 5643 of 
$\mu$ = 30.57 $\pm$ 0.05 mag and for M101 of
$\mu$ = 29.194$\pm$0.039 mag in Riess et al.  (2022) had 
central larger values  than those from the TRGB, despite the higher H$_{0}$ around 73 km s$^{-1}$ Mpc $^{-1}$
obtained by this method compared with 69 km s$^{-1}$ Mpc$^{-1}$
favored by the TRGB.
This could have been a fluctuation. However, the new values published in 2024
put back in agreement the distances for those two galaxies  with those of the CCHP (Freedman et al. 2019) using the TRGB method.

\section{Discussion}

\bigskip

\noindent
While the use of twins has been proved to be very useful for distance determinations,  pairing SNe Ia which are not true twins 
can lead
to  errors in distance estimates. A good first order guide is to
explore only SNe Ia within similar stretch values s$_{BV}$. Then, within a
similar stretch family, one can compare the spectra at various phases, as we
have done here. 
We have looked to twins of the SN 2013aa/SN 2017cbv class, as well as to the
SN 2011fe class.

\bigskip

\noindent
How to size the difference in between SNe Ia twin classes?.
If we placed  2013aa at the distance of 6.5 Mpc for
SN 2011fe, we find a factor of 1.46 of difference in flux level. If we
wrongly took SN 2013aa for a twin of SN 2011fe, the error in distance
would be by a factor of 1.208 (see Figure 11). 

\begin{figure}
  \includegraphics[width=0.9\textwidth]{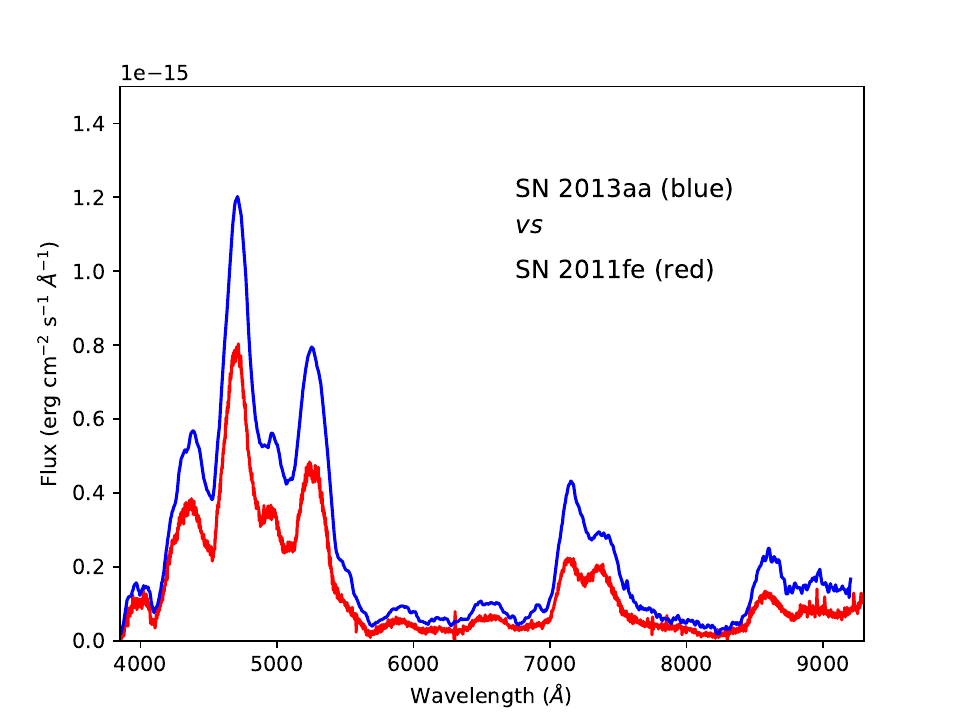}
\caption{Comparison of the spectra of SN 2013aa and SN 2011fe at the same
  phase. The reddening of SN 2011fe has been placed at E(B-V)$=$ 0.15 to
compare it with SN2013aa.}
\end{figure}

\bigskip

\noindent
There is  a lack of the spectral line
of stable Ni in SN 2013aa, which
indicates that the WD ignited at a lower
central density than in SN 2011fe there. The line due to the
presence of stable Ni isotopes points to a start of the ignition 
at high densities, in SN 2011fe.
Both SN 2011fe and SN 2018gv underwent
explosive burning at a higher density
than SN 2017cbv and SN 2013aa.
The mismatch between them is seen at both early and late phases.
In fact,
SN 2013aa and SN 2017cbv synthesized about 0.23 M$_{\odot}$ more  $^{56}$Ni than
the pair SN 2011fe and SN 2018gv, as suggested by the overall higher flux. 
This is corroborated by previous studies
(Jacobson-Gal\'an et al. 2018) suggesting 
0.732 $\pm$ 0.151 M$_{\odot}$ of $^{56}$Ni  in SN 2013aa.
Those authors obtained such a value for $^{56}$Ni in SN 2013aa 
using  a particular approach at late phases (a sort of rate of decline of the
bolometric
light curves, with stretch described in Graur et al. 2018) and using as well
the Arnett's law at early phases.

\bigskip

\noindent
It seems that the class of SN 2013aa/SN 2017cbv is well understood and
if a SNIa of this kind would happen in galaxies far enough to have a
measured redshift not affected by peculiar velocities, a good
measurement of H$_{0}$ should be possible.

\subsection{The whole SNe Ia calibrating sample}

\bigskip

\noindent
By looking at the nebular spectra of the sample of SNe Ia in the calibrating
samples of galaxies of 
TRGB  (18 SNe Ia in 15 galaxies) of Freedman et al. (2019), and in the
calibrating sample of 42 SNe Ia in 37 galaxy hosts of Riess et al. (2022), 
we have
twins of SN 2011fe, as well as twins of
SN 2017cbv/2013aa.
There is also the class of SNe Ia with light echos, which can be close
to the normal SN 2011fe or to the overluminous  SN 1991T in those calibrating
samples.
In fact, SN 1991T had an echo. There seems
to be
as well a class represented by SN 2012cg with its twin SN 2009ig,
which show evidence of interaction with a non-degenerate companion.
The lack of strong lines of
stable Ni points to ignition at a lower density than in SN 2011fe, there.

\bigskip

\noindent
In general, in those SNe Ia lists, 
there is a range of SNe Ia which have synthesized different amounts
of $^{56}$Ni (from 0.4 to 0.8 M$_{\odot}$) and have undergone thermonuclear
burning at different
densities. In addition, SNe Ia do have peculiarities such as evidence for an
early
interaction with a supernova companion, or the presence of an echo at
late times.

\bigskip

\noindent
For the first time, a long coverage of a SN Ia  at the nebular phase
has been obtained, from 400 days to 1000 days in the case of SN 2011fe
(Tucker et al 2022). We have seen clearly how the ionization stage changes
significantly along this time, for this normal SN Ia. But even from 200 to
460 days the level of ionization of Fe$^{+}$ and Fe$^{++}$ undergoes
significant change.

\bigskip
\noindent
A dense grid of nebular spectra of
SNe Ia would facilitate the twins classification. The use of this
database should enable 
comparison of spectra of different supernovae at similar dates. The 
use of different SNe Ia at significant different phases, as if they were at the same phase, which is nowadays often done at late  times, 
should be avoided. It affects the
evaluation of the masses of NSE nuclides synthesized in the
explosion. The abundances of stable NSE nuclides are indicators of
the densities at which the WDs
explode and they also give clues about the mechanism involved in the explosion.
But, for the present case, the overall flux of the elements from the
decay of $^{56}$Ni synthesized in the explosion make the overall flux of
the spectra at late times, therefore affecting any distance determination. 

\bigskip

\noindent
 Reliable distances can be obtained by pairing the twins until the
nebular phase, once proper flux calibration and choice of similar
phases are done. Including the
nebular phase can prevent errors made when matching the very early phases,
at which times different interactions with circumstellar
material or possible heating from a companion
would produce differences in the spectra.

\bigskip

\noindent
The distance to M101 from Beaton et al. (2019) using
the TRGB method is consistent  with SN 2011fe having 
synthesized 0.5 M$_{\odot}$ of $^{56}$Ni, as most authors have found. 

\bigskip

\noindent
 The difference between SN 2011fe and the SN 2017cbv/2013aa suggests
that that NGC 5643 is at $\mu$ = 30.48 $\pm$ 0.065 mag. The convergent value on M101 from previous section is $\mu$ $=$ 29.075 $\pm$ 0.068 mag and 
NGC 7250 is  at $\Delta d$ = 7.64 $\pm$ 0.03 Mpc from our joined twin
SNe Ia spectra from NGC 5643, thus at 
 31.518 $\pm$ 0.065 mag.
  The NGC 5643 value  derived by the Cepheids is 
31.628 $\pm$ 0.126 mag from {\it SH0ES} (Riess et al. 2022), which does not yet
use {\it JWST} and would likely be reexamined. 

\bigskip

\noindent
     We include in Table 8
      the distance values using the ''SNe Ia twins for life''
      method as well as those provided by other methods, as discussed
      in the main text.

\begin{table}[ht!]
\centering
\caption{Moduli and distances from the present work {\it vs} Cepheid and
  TRGB moduli}

\begin{tabular}{lcccc}
  \\
    \hline
    Galaxy & $\mu$ (mag) & D (Mpc) & $\mu$ Cepheids (mag)$^{1}$ & $\mu$ TRGB (mag)  \\
    \hline
    \hline \\
    M 101     & 29.075$\pm$0.068 & 6.53$\pm$0.20 &
    29.194$\pm$0.039$^{2}$ & 29.08$\pm$0.04$^{3}$ \\ \\
    NGC 5643  & 30.48$\pm$ 0.065 & 12.47$\pm$ 0.37  &
    30.570$\pm$0.050$^{4}$ & 30.48$\pm$ 0.1$^{5}$ \\ \\
NGC 7250  & 31.518$\pm$0.065 & 20.12$\pm$0.70& 31.628$\pm$0.126 &  $\dots$ \\ \\
NGC 2525  & 31.933$\pm$0.069 & 24.35$\pm$0.77 & 32.067$\pm$0.100 &
$\dots$\\ \\
\hline

\end{tabular}

$^{1}$Riess et al. (2022). $^{2}$29.10$\pm$0.06 from Mira variables (Huang et al. 2024). $^{3}$Freedman et al. (2019). $^{4}$Changed to
30.52$\pm$0.02 in Riess et al. (2024a). $^{5}$Hoyt et al. (2021).
\end{table}

\bigskip

\noindent
The method has been applied as well to the twins of SN 2011fe in M101 and
SN 2018gv in NGC 2525. This has allowed to determine a distance of
 24.35$\pm$0.77 Mpc 
 to NGC 2525. The value reported by Riess et
al. (2022) is 25.90$\pm$1 Mpc. Our result presented here is
compatible with that from the Cepheids due to the large error on the
  distance determination. It does not
align well with the  extreme distance  $\sim$ 27 Mpc in the upper edge of
 the uncertainty towards NGC 2525 obtained from the
Cepheids. 
The {\it SH0ES} collaboration measurement places SN 2018gv at a relative
distance factor $\sim$
3.75 times that of SN 2011fe, while we favor a factor
$\sim$ 3.71.\footnote{
  Very recently, new measurements of the  distances to the
  galaxies in Table 8 have been published. They have been obtained with
  data from the {\it JWST} by Freedman et al. (2024). For the galaxy
  NGC 7250, they obtain using Cepheids a distance modulus of
  31.41 $\pm$0.12 mag 
   while using TRGB they find a distance of 31.62 $\pm$ 0.04 mag.
   So, our measurement lies in the middle, being compatible within 1$\sigma$.
   This team has revised as well the distance to NGC 5643 with the new
   {\it JWST} data and they find a distance modulus of 30.51 $\pm$ 0.08 mag
   using Cepheids and 30.61 $\pm$ 0.07 mag using TRGBs. Our favored distance
   is within 1$\sigma$. The {\it SH0ES} team has, on the other hand,
   revised their distance to NGC 2525 using TRGBs and they find
   a distance modulus of 31.81 $\pm$ 0.80 mag, compatible with our distance, but
   now with a smaller error bar and lower central value for the distance
   modulus, as we had suggested. }

\bigskip

\noindent
It seems clear that twins followed until the nebular phase can establish the distance
ladder in the epoch of large surveys. 
The infrared part of the nebular spectra of SNe Ia will soon become widely
available, from 
programs already approved for the {\it JWST}. The infrared will reveal further
differences
among SNe Ia in that part of the spectra, thus helping in the twin
classification. 

\section{CONCLUSIONS}

\bigskip

\noindent
It has been proved here that comparison of twin SNe Ia  can provide a robust
way to establish the extragalactic distance ladder. The method has
been applied to the twins in the galaxy NGC 5643: SN 2013aa and
SN 2017cbv. The comparison using spectra before
maximum and at the nebular phase shows that the error in the distance
determination is of $\Delta \mu$ $\sim$ 0.005 mag. 

\bigskip

\noindent
Distances obtained here for NGC 7250 and NGC 2525 are consistent with those
obtained by Riess et al. (2022) given the large error bars of the distances to
these galaxies by those authors. Our distances are somehow shorter
and with smaller errors than those
in the mentioned paper. Now, the very  recently  revised
value published
in Li et al. (2024) to NGC 2525 agrees with our suggestion made since the
first version of this manuscript. 
 The value of the distance to
NGC 5643 (Hoyt et al. 2021) is in good accordance with the one derived
by the SNe Ia in this paper, but not with that  was presented by 
Riess et al. (2022).  The difference 
of  $\mu_{Cepheids}$ $=$ 30.57 $\pm$ 0.05 mag versus $\mu_{TRGB}$ $=$ 30.48$\pm$0.1
mag by Hoyt et al. (2021) is
larger than the usual 0.05 mag ( $\Delta \mu$ $\sim$ 0.1 ), leaving aside
the errors. If one takes into account that NGC 5643 is a nearby galaxy, the
difference called for a revision, as we mentioned in an earlier version of
this paper. Such revision has been
  done in Riess et al. (2024b) with Cepheids in NGC 5643 observed with
  the {\it JWST} and a value of $\mu$ $=$ 30.52 $\pm$ 0.02 has been
  obtained providing a much better agreement between methods. We found a need of a similar revision  of the Cepheids distance to NGC 7250 and NGC 2525
   in the
    first version of this manuscript, and this revision has been
    made successfully. Our method has proved to find  inconsistencies
    in the distance moduli given by Cepheids or TRGB, therefore being very
     useful.  More distance moduli will soon see the light
    from the {\it JWST} data analysis.

\bigskip

\noindent
In the near future, we plan
to complete the check for the galaxy sample measured by Cepheids and
TRGB published so far by applying our method.
But we want to go to distant galaxies
as a priority, as we suspect that a major difference in the H$_{0}$ value from
Cepheids and from TRGBs might arise from the different procedure followed to
link the local SNe Ia sample with that one in the Hubble flow.

\bigskip 

\noindent
The way to  evade the problem is to obtain H$_{0}$ 
using SNe Ia twins in galaxies of the Coma cluster or any galaxy at z $\sim$ 0.02--0.03, already in
the Hubble flow and compare them with their local SNe Ia twins. 
We would like to extend the method of the twin distance determinations to
SNe Ia in the Hubble flow  observing spectra not only near maximum, but
also well into the epochs where SNe Ia show the diversity of the inner core,
specially the Fe-peak elements and $^{56}$Ni--rich innermost layers. That
should be achievable with the {\it JWST} or the {\it ELT}. 

\bigskip

\bigskip

\noindent
The authors would like to thank Melissa Graham and Chris Burns for generously
providing spectra used in the present paper.
This work has made use as well of spectra from the {\it Weizmann Interactive
  Supernova data REPository} ({\it WISeREP}).
Gaia DR3 photometry was used in the calibration of SN 2018gv/Gaia18bat data.
We acknowledge {\it ESA Gaia, DPAC and the Photometric Science Alerts
Team (http://gsaweb.ast.cam.ac.uk/alerts)}.
PR-L would like to thank
Michael Weiler and Josep Manel Carrasco from the Gaia--ICCUB team for their help
in crosschecking the calibration
of the spectra of SN 2018gv at late phases. 
PR-L also acknowledges support from grant PID2021-123528NB-I00, from the
the Spanish Ministry of Science and Innovation
(MICINN).
JIGH acknowledges financial
support from MICINN grant   
PID2020-117493GB-I00.

\bigskip

\bigskip

\noindent
{\bf REFERENCES}

  \bigskip

 \noindent
Anand, G.A., Riess, A.G., Yuan, W., et al. 2024, ApJ, 996, 89

\bigskip

\noindent
  Beaton, R.L., Seibert, M., Hatt, D., et al. 2019, ApJ, 885, 141

  \bigskip

 \noindent
  Bernal, J.L., Verde, L., \& Riess, A.G. 2016, JCAP, 10, 019

  \bigskip

  \noindent
  Blakeslee, J.P., Jensen, J.B., Ma, C.-P., Milne, P.A., \& Greene, J.E. 2021,
  ApJ, 911, 65

  \bigskip

  \noindent
  Bloom, J.S., Kasen, D., Sen, K.J., et al. 2012, ApJL, 144, L17

  \bigskip

  \noindent
  Boone, K., Fakhouri, H., Aldering, G.A., et al. 2016, AAS Meeting, 227,
  237.10, NASA ADS: 2016AAS…22723710B

  \bigskip

\noindent
Boone, K., Aldering, G., Antilogus, P., et al. 2021, ApJ, 912, 71

\bigskip

\noindent
Branch, D., Dang, L.C., Hall, N., et al. 2006, PASP, 118, 560
  
\bigskip

\noindent
Brown, P.J., Dawson, K.S., Harris, D.W., eyt al. 2012, ApJ, 749, 18

\bigskip

\noindent
Burns, C.R., Stritzinger, M., Phillips, M.M., et al. 2014, ApJ, 789, 32 

\bigskip

\noindent
Burns, C.R., Ashall, C., Contreras, C., et al. 2020, ApJ, 895, 118

\bigskip

\noindent
  Casper, C., Zheng, W., Li, W., et al. 2013, CBET, 3588, 1,
       NASA ADS: 2013CBET.3588....1C

       \bigskip

 \noindent      
  Cenko, S.B., Thomas, R.C., Nugent, P.E., et al. 2011, ATel, 3583, 1

  \bigskip

  \noindent
  de Vaucouleurs, G., de Vaucouleurs, A., Corwin, H.G. et al. 1991,
  Third Reference Catalogue of Bright Galaxies,
  Vol. 1 (Springer, New York), ISBN  3-540-97552-7; 0-387-97552-7

  \bigskip

  \noindent
  Di Valentino, E., Melchiorri, A., \& Silk, J. 2021, ApJL, 908, L9

  \bigskip

  \noindent
    Efstathiou, G., Rosenberg, E., \& Poulin, V. 2024, Phys. Rev. Lett.,
     132, 221002

     \bigskip

\noindent     
Fakhouri, H.K., Boone, K., Aldering, G., et al. 2015, ApJ, 815, 58

\bigskip

\noindent
  Foreman-Mackey, D., Hoog, D.W., Lang, D., \& Gooodman, J. 2013, PASP, 125,
  300

  \bigskip

  \noindent
    Freedman, W.L., Madore, B.F., Hatt, D., et al. 2019, ApJ, 882, 34

    \bigskip

  \noindent
  Freedman, W.L., \& Madore, B.F. 2023, JCAP, 11, 050

  \bigskip

\noindent  
Freedman, W.L., Madore, B.F., Jiang, I.S., Hoyt, T.J., Lee, A.J., \& Owens, K.A. 2024, arXiv:2408.06153
  
\bigskip

\noindent
    Gelman, A., Carlin, J.B., Stern, H.S., Dunson, D.B., Vehtari,
    A., \& Rubin, D.B. 2014,
     Bayesian Data Analysis, by A. Gelman et al. Third edition, Boca Raton, FL,
     Chapman \& Hall, ISBN 13-9781-43984-0955, NASA ADS: 2014bda..book.....G

     \bigskip

 \noindent
  Goodman, J., \& Weare, J. 2010, Communications in Applied Mathematics and
  Computational Science, 5, 65,  DOI: 10.2140/camcos.2010.5.65

  \bigskip

  \noindent
  Graham, M.L., Kennedy, T.D., Kumar, S., et al. 2022, MNRAS, 511, 3682

  \bigskip

 \noindent
Graur, Or, Zurek, D.R., Cara, M., et al. 2018, ApJ, 866, 10
  
\bigskip

\noindent
  Guy, J., Astier, P., Nobili, S., Regnault, N., \& Pain, R. 2005, A\&A, 443,
  781

  \bigskip

  \noindent
  Hoyt, T., Beaton, R.L., Freedman, W.L., et al. 2021, ApJ, 915, 34 

  \bigskip

\noindent  
Hamuy, M., Phillips, M.M., Maza, J. et al. 1995, AJ, 109, 1

\bigskip

\noindent
Hamuy, M., Cartier, R., Contreras, C., \& Suntzeff, N.B. 2021, MNRAS, 500, 1095

\bigskip

\noindent
Huang, C.D., Yuan, W., Riess, A.G., et al. 2024, ApJ, 963, 83

\bigskip

\noindent
  Itagaki, K. 2018, TNSTR, 57, 1

  \bigskip

\noindent
Jacobson-Galan, V.W., Dimitriadis, G., Foley, R.J., \& Kirpatrick, C.D.
2018, ApJ, 857, 88

\bigskip

\noindent
  Jha, S., Riess, A.G., \& Kirshner, R.P. 2007, ApJ, 659, 122

  \bigskip

\noindent
  Kelson, D.D., Illingworth, G.D., Freedman, W.L., et al. 1996, ApJ, 463, 26

  \bigskip

  \noindent
  Kenworthy, W.D., Riess, A.G., Scolnic, D., et al. 2022, ApJ, 935, 83

  \bigskip

\noindent  
Khetan, N., Izzo, L., Branchesi, M., et al. 2021, A\&A, 647, A72
 
\bigskip

\noindent
  Li, S., Anand, G.S., Riess, A.G., et al. 2024, arXiv:2408.0006

  \bigskip

  \noindent
  Li, W., Bloom, J.S., Podsiadlowski, P., et al. 2011, Natur, 480, 348

\bigskip
  
  \noindent
  Madore, B.F., \& Freedman, W.L. 2024, ApJ,  961, 166

  \bigskip

  \noindent
    Mazzali, P.A., Sullivan, M., Filippenko, A.V., et al. 2015, MNRAS, 450, 2631

    \bigskip

\noindent    
Morrell, N., Phillips, M.M., Folatelli, G., et al. 2024, ApJ, 967, 20
    
\bigskip

\noindent
  Murakami, Y.S., Riess, A.G., Stahl, B.E., et al. 2023, JCAP, 11, 046

  \bigskip

\noindent  
Nomoto, K., Thielemann, F.-K., \& Yokoi, K. 1984, ApJ, 286, 644

\bigskip

\noindent
Nugent, P., Phillips, M.M., Baron, E., Branch, D., \& Hauschildt, P. 1995,
ApJL, 455, L147

\bigskip

\noindent
Nugent, P.E., Sullivan, M., Cenko, S.B., et al. 2011, Natur, 480, 344

\bigskip

\noindent
  Pan, Y.C., Foley, R.J., Kromer, M., et al. 2015, MNRAS, 452, 4307
  
  \bigskip

  \noindent
  Parrent, J.T., Sand, D., Valento, M., Graham, D.A., \& Howell, D.A. 2013,
  ATel, 4817, 1

  \bigskip

  \noindent
    Patat, F., Cordiner, M. A., Cox, N. L. J., et al. 2013, A\&A, 549, A62

    \bigskip

    \noindent
    Pereira, R., Thomas, R.C., Aldering, G., et al. 2013, A\&A, 554, A27

    \bigskip

\noindent    
Peterson, E.R., D'Arcy Kenworthy, W., Scolnic, D., et al. 2022, ApJ, 938, 112

\bigskip

\noindent
  Perlmutter, S., Aldering, G., Goldhaber, G., et al. 1999, ApJ, 517, 565

  \bigskip

  \noindent
  Phillips, M.M. 1993, ApJL, 413, L105

  \bigskip

  \noindent
  Phillips, M.M., Lira, P., Suntzeff, N.B., et el. 1999, AJ, 118, 1766 

  \bigskip

\noindent  
Planck Collaboration 2020, A\&A, 641, A6

\bigskip

\noindent
    Richmond, M.W., \& Smith, H.A. 2012, JAVSO, 40, 872

    \bigskip

 \noindent   
    Riess, A.G., Press, W.H., \& Kirshner, R.P. 1996, ApJ, 473, 88

    \bigskip

  \noindent  
  Riess, A.G., Filippenko, A.V., Challis, P., et al. 1998, AJ, 116, 1009

  \bigskip

  \noindent
    Riess, A.G., Macri, L.M., Hoffmann, S.L., et al. 2016, ApJ, 826, 56

    \bigskip

 \noindent
  Riess, A.G., Yuan, W., Macri, L.M., et al. 2022, ApJ, 934, L7

  \bigskip

  \noindent
  Riess, A.G., Anand, G.S., Yuan, W., et al. 2024a, ApJL, 962, L17

  \bigskip

  \noindent
 Riess, A.G., Scolnic, D.,  Anand, G.S., et al. 2024b, arXiv:2408.11770

 \bigskip

 \noindent
  Rubin, D., Aldering, G., Betoule, M., et al. 2023, arXiv:2311.12098
                                                                     
  \bigskip

 \noindent
Ruiz--Lapuente, P., \& Lucy, L.B., 1992 ApJ, 400, 127

\bigskip

\noindent
    Ruiz--Lapuente, P. 1996, ApJ, 465, L83

    \bigskip

    \noindent
   Shappee, B.J., \& Stanek, K.Z. 2011, ApJ, 733, 124 

   \bigskip

   \noindent
Schlafly, E.F., \& Finkbeiner, D.P. 2011, ApJ, 737, 103

\bigskip

\noindent
    Schneider, S.E., Thuan, T.X., Mangum, J.G., \& Miller, J. 1992, ApJS, 81, 5

    \bigskip

    \noindent
    Scolnic, D., Brout, D., Carr, A., et al. 2022, ApJ, 938, 113
    
    \bigskip

    \noindent
  Shingles, L.J., Sim, S.A., Kromer, M., et al. 2020, MNRAS, 492, 2029

  \bigskip

  \noindent
  Shingles, L.J., Flors, A., Sim, S.A, et al. 2022, MNRAS, 512, 6150

  \bigskip

  \noindent
  Stritzinger, M., Hamuy, M., Suntzeff, N.B., et al. 2002, AJ, 124, 2100

  \bigskip

  \noindent
  Suntzeff, N.B., Hamuy, M., Martin, G., Goomez, A., \& Gonzalez, R. 1988,
  AJ, 96, 1864

  \bigskip

  \noindent
  Tammann, G.A., \& Reindl, B. 2011, arXiv:1112.0439

  \bigskip

  \noindent
  Tartaglia, R., Sand, D., Wyatt, S., et al. 2017, 10158,
     NASA ADS: ATel10158....1T

     \bigskip

 \noindent
  Tucker, M.A., Shappee, B.J., Kochanek, C.S., et al. 2022, MNRAS, 517, 4119

  \bigskip

  \noindent
  Tucker, M.A., \& Shappee, B.J. 2024, ApJ, 962, 74

  \bigskip

  \noindent
  Waagen, E.O. 2011, AAN, 446, 1

  \bigskip

  \noindent
  Waagen, E.O. 2013, AAN, 479, 1

  \bigskip

  \noindent
  Wilk, K.D., Hiller, D.J., \& Dessart, L. 2018, MNRAS, 474, 3187

  \bigskip

  \noindent
  Yang, Y., Hoeflich, P., Baade, D., et al. 2020, ApJ, 902, 46

  \bigskip

  \noindent
  Zhai, Q., Zhang, J.-J., Wang, X.F., et al. 2016, AJ, 151, 125

  \bigskip

  \noindent
Zhang, K., Wang, X., Zhang, J., et al. 2016, ApJ, 820, 677

\bigskip

\noindent
    Zheng, W., Silverman, J.M., Filippenko, A.V., et al. 2013, ApJL, 788, L15

\clearpage

\noindent
    {\bf APPENDIX}

\bigskip

\noindent
    {\bf Appendix A. Signatures for identifying  twins.
      The example of SN 2017cbv/2013aa and SN 2011fe}

    \bigskip

    \noindent
    We have stated that twins belong to a similar stretch class and, moreover,
    they only differ in colours by around 0.04 mag.
    The shape of the pseudo--continuum should be similar, therefore.
    We address this comparison in the first place. 
    One can also expect that the
    spectral features in twin SNe Ia will be very similar.
    This can be quantified by the usual ratios that characterize
    the sample of SNe Ia. We refer here, in the second place,
    to the work on pseudo--equivalent
    widths among SNe Ia. In the third place, we will refer to the ratio of
    the two lines of Si II at $\lambda$ 5972 \AA \ and
    $\lambda$ 6355 \AA .

    \bigskip

\noindent 
{\bf I. Pseudo--continuum.}

\bigskip

\noindent
In the early phases the amount of $^{56}$Ni synthesized in the SNIa explosion
determines to have a hotter or cooler photosphere. A SN Ia with
more $^{56}$Ni will show a
higher effective temperature (T$_{\rm eff}$) of the pseudo-continuum.

\bigskip

\noindent
We have compared the spectra of SN 2017cbv/SN2013aa at -2d with that of
SN 2011fe and it shows that the shape of the pseudo--continuum is that
of a Planck function with T$_{\rm eff}$ of 25,000 K while for the same
phase in SN 2011fe the shape provides 23,000 K, thus 2000 K lower
(see Figure 12).
We have also compared the spectrum of SN 2017cbv at 12 days past maximum and
the difference is also of 2000K. The spectrum of SN 2017cbv shows a
pseudo--continuum of 21,000 K while the spectrum of SN 2011fe fits
with a T$_{\rm eff}$  of 19,000 K (see Figure 13). 

\begin{figure}[H]
\includegraphics[width=0.75\textwidth]{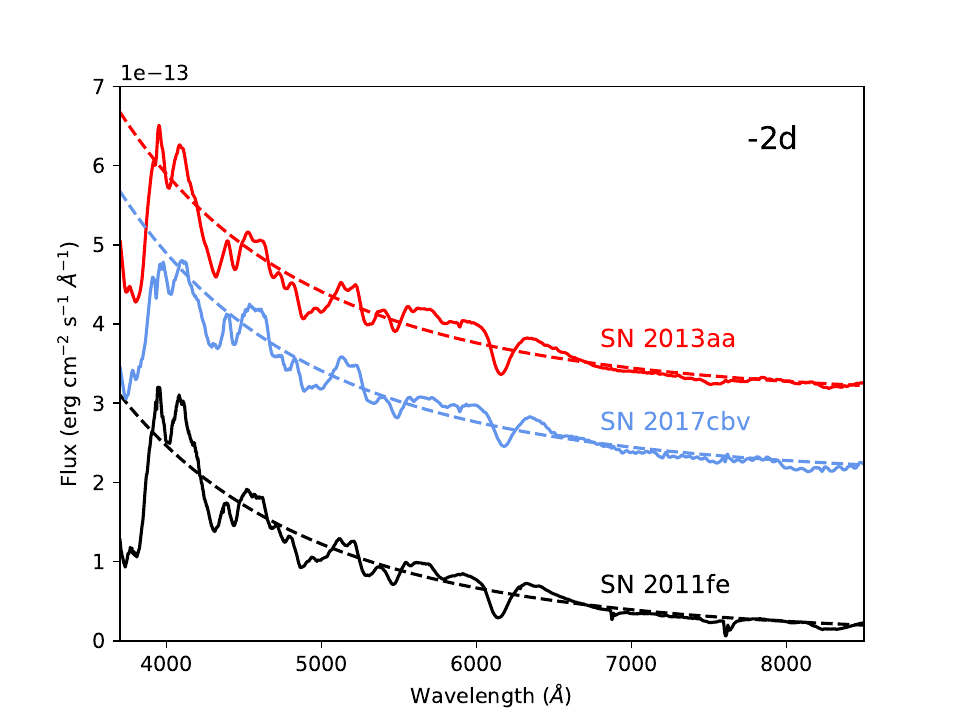}  
\caption{Comparison of SN 2017cbv/SN2013aa with SN2011fe at 2 days before
    maximum. The T$_{\rm eff}$ of the photosphere of SN 2017cbv/2013aa is 2000 K
    higher than in SN 2011fe.}
\end{figure}

\begin{figure}[H]
\includegraphics[width=0.75\textwidth]{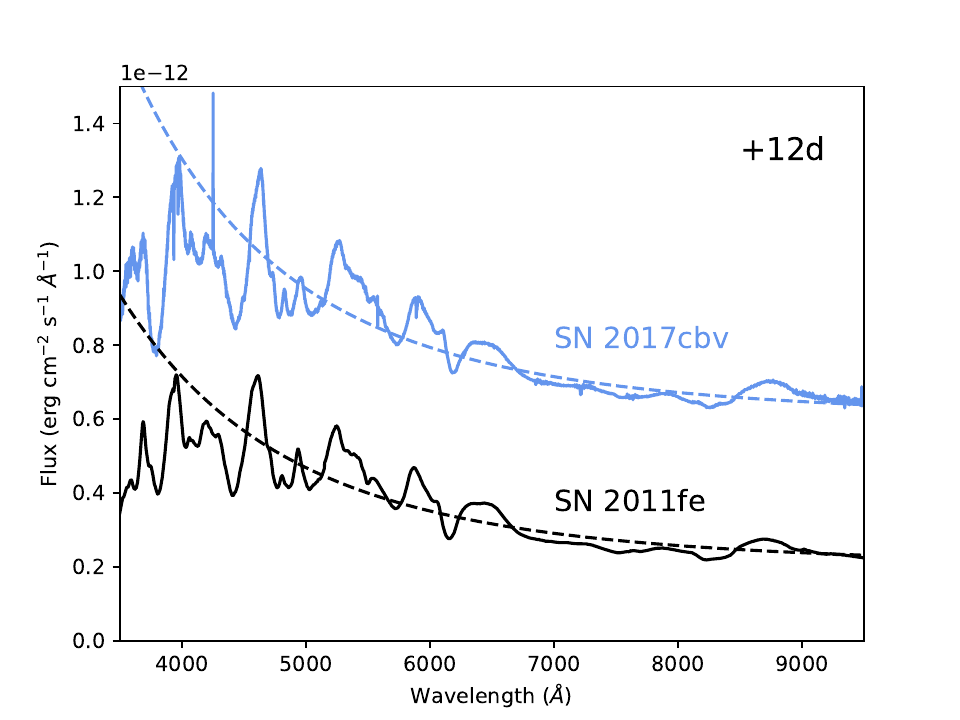}  
\caption{Comparison of SN 2017cbv/SN2013aa with SN2011fe at 12 days past
    maximum. The T$_{\rm eff}$ of the photosphere of SN 2017cbv is 2000 K
   higher than in SN 2011fe. }
\end{figure}

\bigskip

    \noindent
        {\bf II. pWs}

        \bigskip

        \noindent
        The pseudo--equivalent widths of different lines near maximum of
        SNe Ia are the basis 
        of  the Branch classification of SNe Ia, which
        informs about the inner distributions of elements in
        the velocity space of the supernova ejecta. The classes are ''shallow
        silicon'' (SS), ''core normal'' (CN), ''broad line (BL) and
        ''cool line'' (CL) (Branch et al, 2006). SN 2017cbv and SN 2013aa
        belong to the ''core normal'' class, as SN 2011fe. Burns et al. (2020)
        measure the pseudo--equivalent widths of SN 2017cbv and SN 2013aa
        and, in their Figure 8,  pWs of $\lambda$ 5972 \AA \ versus  
        $\lambda$ 6355 \AA \ reveal that both SNe Ia are
        very close in the core--normal
        region of the Branch diagram.  We have measured the pWs of those
        lines and found that they are as in Burns et al. (2020).
        SN 2011fe is also a CN SNe Ia, but shows different pWs than the class
        of SN 2017cbv and SN 2013aa.

\bigskip

\noindent
   The depths of the various lines is a measure of their similarity
   as SNe Ia explosions and this is correlated with the stretch/rate
   of decline s$_{BV}$/ $\Delta m_{15}$ (see, for the latest on this,
   Morrell et al. 2024).

   \bigskip
   
   \noindent
   The diagnostics is clearer for the SNe Ia spectra which do not come with
   blends from other lines. This is the case for CaII H \& K (pW1), and 
    the double S II ''W' feature
   at $~$   5400 \AA \, which 
   is a clear blend  dominated by S II (pW5). It is also the case for the
   Si II $\lambda$ 5972 \AA \ line (pW6) and the
   Si II $\lambda$ 6355 \AA (pW7).  We can see the difference
   of the values between SN 2017cbv/2013aa and SN 2011fe (see Table 9).
   The trend
   between the Si II pWs of the mentioned lines
   and the rate of decline s$_{BV}$ or m$(B)_{15}$ is well illustrated in
   Morrell et al. (2024). We confirm such correlation. 

   \bigskip

   \begin{table}[h!]

  \footnotesize

  \centering
  \caption{Photometric and spectroscopic characteristics of SNe Ia}

  \begin{tabular}{lcccccc}
    \hline
    Supernova & $\Delta m_{15}$ (B) & $s^{D}_{BV}$ & $pW1$ (Ca II H\&K) &
    $pW5$ (S II W) & $pW6$ (Si II 5972) & $pW7$ (Si II 6355)  \\
    &  [mag]  &     & [\AA] & [\AA] & [\AA] & [\AA] \\
    \hline
    \hline
    SN 2011fe & 1.07$\pm$0.06 & 0.919$\pm$0.004 & 111$\pm$1 & 74$\pm$1 &
    16.0$\pm$0.5 & 98$\pm$1 \\
    SN 2013aa &  0.96$\pm$0.01 & 1.11$\pm$0.02 & 74$\pm$1 & 67$\pm$1 &
    10.0$\pm$0.5 & 84$\pm$1 \\ 
    SN 2017cbv & 0.96$\pm$0.02 & 1.11$\pm$0.03 & 64$\pm$1 & 69$\pm$1 &
    10.0$\pm$0.5 & 76$\pm$1 \\
    \hline

\end{tabular}
\end{table}

\bigskip
   
\noindent
    {\bf III. Si II ratio  $\Re_{Si}$}

    \bigskip

\begin{table}[h!]
  \centering
  \caption{Si II $\lambda\lambda$5972/6355 ratios}
  \begin{tabular}{lc}
    \hline
    Supernova & $\Re_{Si}$   \\
    \hline
    \hline
    SN 2011fe & 0.23$\pm$0.05  \\
    SN 2013aa & 0.10$\pm$0.05  \\
    SN 2017cbv & 0.13$\pm$0.05 \\  
    \hline
\end{tabular}
\end{table}

\bigskip

\noindent
   The ratio of the two Si II lines, Si II $\lambda$ 6355 \AA \
   and Si II  $\lambda$ 5972 \AA \
   defines the $\Re_{Si II}$
   parameter as introduced by Nugent et al. (1995).
   According
   to Nugent et al (1995), the Si II lines interact with line blanketing
   from Fe III and Co III
at pre--maximum when the temperature is high and Fe and Co are substantially present in
the outer layers. This effect washes out the $\lambda$ 5972 \AA \ line and makes
de $\Re_{Si}$ lower than for SNe Ia with lower temperatures near and in the 
pre--maximum. Here we show how in SN 2017cbv and SN 2013aa  $\Re_{Si}$ is lower
than for SN 2011fe (see Table 10).  We underlined in the first
subsection of this Appendix A
how the effective
temperatures of SN 2017cbv and SN 2013aa are typically 2000K higher than in SN 2011fe at
various phases. So, this is consistent with the different $\Re_{Si II}$.

\bigskip

    \noindent
    {\bf B. Theoretical anchor for SN 2011fe in M101}

    \bigskip

\begin{figure}[ht!]
  \centering
\includegraphics[width=1.0\textwidth]{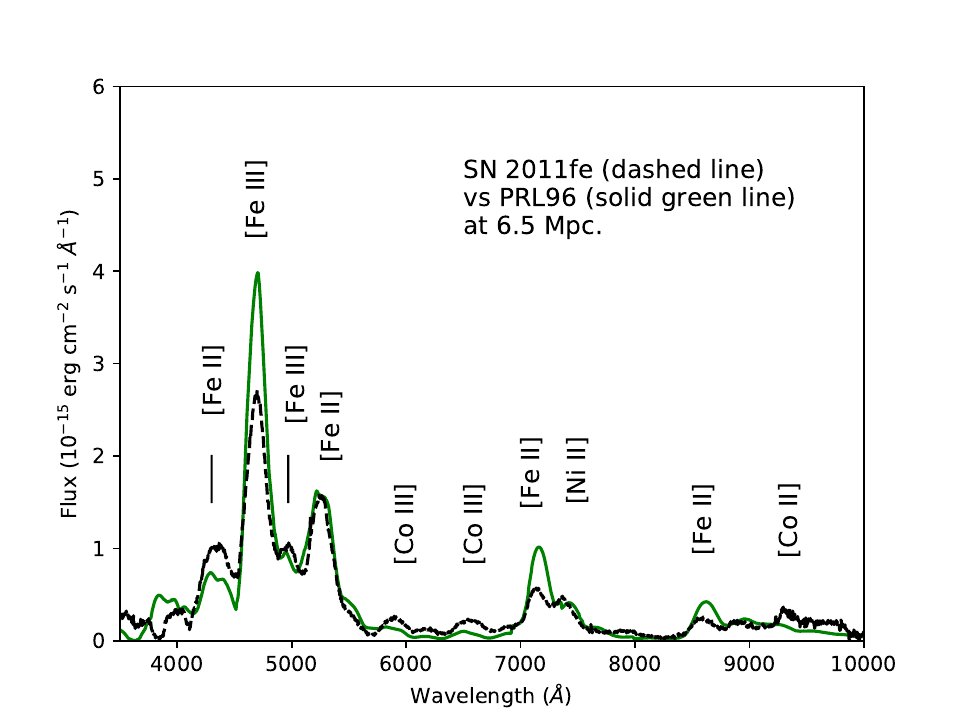}
\caption{A distance of 6.5 Mpc seems compatible with the late time spectrum of
  SN 2011fe. Here a comparison is made with a theoretical spectrum with a
 $^{56}$Ni mass of 0.5 M$\odot$ of the W7 class obtained with the code explained in Ruiz--Lapuente (1996).}
  \end{figure}

\bigskip

\noindent
 At late phases, the
population of the energy levels of the ions present in the supernova ejecta is
out of LTE because the density of the electrons responsible for collisionally
exciting the lines is lower than the critical
density for the corresponding transitions. Forbidden emission can be exploited
here, since such emission from iron ions (especially Fe$^{+}$) trace very well
  the electron density
  (n$_{e}$) of the supernova ejecta ``nebula''. This fine sensor of the density
  profile can help to probe the mass of the exploding white dwarf and
  its kinetic
  energy. A clear diagnostic of low n$_{e}$ comes from the emission at
  $\lambda$5200 \AA, $\lambda$4300 \AA, and $\lambda$5000 \AA. These emissions
  are due to Fe$^{+}$ a$^{4}$F-b$^{4}$P and a$^{4}$F-a$^{4}$H transitions,
  whose lower energy terms can be significantly depopulated if n$_{e}$ is low.
  This ratio becomes a diagnostic of ejected mass. If lower masses give on
  average lower collisional excitation rates for
  those forbidden transitions, the rates decrease significantly.
  Another interesting ratio
  of [Fe II]
  is that of a$^{4}$F-b$^{4}$P to a$^{4}$F-a$^{4}$P, with emissions at
  $\lambda$5262 and $\lambda$8617 \AA, though there are
  emissions at around 8600 \AA \ from other elements that can prevent the
  discrimination. In any case, these [Fe II] lines at long wavelenghth do
  inform on how reddened the SN Ia is. They provide an internal estimate of
  reddening.

  \bigskip

  \noindent
  The second aspect relative to the amount of $^{56}$Ni is that it impacts the
  degree of ionization and electron temperature T$_{e}$ of the ejecta, which
  are kept warm by the thermalization of the $\gamma$-rays and positrons
  (e$^{+}$) from the decay of $^{56}$Co (coming from $^{56}$Ni) into $^{56}$Fe.
  While the
  emissivity of the different forbidden transitions of Fe$^{+}$ keeps track of
  the electron density profile of the supernova, the ratio Fe$^{+}$ to
  Fe$^{++}$ gives information on T$_{e}$ and the ionization of the ejecta.
  Since there is a large number of forbidden lines over
the wavelength range at which SNe Ia are observed, there is also complementary
information to obtain an internal estimate of the reddening. A good description
of the ionization degree of the ejecta depends on the ionization treatment in
the radiation transport code, and on factors that need evaluation, such as
trapping of the positron energy, among others. A
feature inherent to the density of the ejecta is that the ionization stage is
kept high (more Fe III in detriment of Fe II) when the density is low and the
recombination rates decrease.

\bigskip

 \noindent
    A fit with a theoretical spectrum is shown in Figure 14 to  illustrate 
    the ions present and dominating the emission of SN 2011fe at a late
    phase.

\bigskip   

\noindent
There is a
general
agreement, among different authors using different codes, that
sub-Chandra explosion models tend to produce overionized ejecta (Ruiz-Lapuente
1996; Mazzali et al. 2015; Wilk et al 2018, Shingles et al. 2020), though
the more recent  denser sub--Chandrasekhar models seeem to fit as well
as Chandrasekhar
models the spectra of normal SNe Ia. 
In general,  there is an overionization in
    the theoretical
    spectra  calculated for all the models with the available codes and 
    that is being currently  addressed (see Shingles et al. 2022). 
    An internal test at
    the nebular phase showed that the R96 code and the ARTIS code gave 
    the same overall flux level of the nebular spectra for W7 model amongst
    others.

    \bigskip

    \noindent
    Due to the limitations to achieve a perfect fit, and the need
    of being extremely precise in the distance derived with spectra, a decision
    to proceed towards a purely empirical approach was made. This has proved to
    be very successful.

\bigskip

\bigskip

\noindent
    {\bf B.  Historical considerations}

    \bigskip

\noindent
In the 90’s, the use of theoretical models to infer basic properties of the
SNe Ia from spectra taken at late times was proposed (see Ruiz–Lapuente 1996;
Ruiz–Lapuente \& Lucy 1992). The idea was to obtain
at the same time the $^{56}$Ni mass, the reddening and the distance to the SN.
Using this method, Ruiz Lapuente (1996) derived, for a reduced sample of
SNe Ia, a value of the Hubble constant of H$_{0}$ = 68 $\pm$ 6 (statistical)
$\pm$ 7 (systematic) km s$^{-1}$ Mpc$^{-1}$. 
The procedure was first to measure distances to the SNe Ia, then obtain
their absolute B magnitude at peak brightness and 
derive a fiducial absolute magnitude M$_{B}$ for the SNe Ia sample.
Then, this one was tied to a H$_{0}$ value as in Hamuy et al. (1995):

\begin{equation}
  {\rm log} H_{0}=0.2\{M^{B}_{\rm MAX} -1.624(\pm0.582)[\Delta m_{15}(B)-1.1]
   +28.296(\pm0.080)\}
\end{equation}

\bigskip

\noindent
 Such calibration has changed dramatically since the 90's. Now the
 {\it Phillips relation} includes terms in higher orders of
 $[\Delta m_{15}(B)-1.1]$. The
  derivation of H$_{0}$ relies
 on the connection between the second and third rung in the cosmic
 distance ladder, i.e from SNe Ia which are not in the
 Hubble flow  (with z well below 0.1)  and those which are already
 in the Hubble flow. 
 The process used by most collaborations still relies in   
 obtaining a  fiducial absolute magnitude in B of the SNe Ia
 at peak, M$^{0}_{B}$,  in the local SNe Ia sample. However, such derivation
 needs corrections from
 extinction, variations of colors at peak maximum brightness in the sample,
 dependency with the host galaxy. In addition. to place them in the
 Hubble flow  a reliable intercept with the SNe Ia in
 the Hubble flow is required. Given the complexity of this route, it is
 not strange to find that M$^{0}_{B}$ varies amongst publications.

\noindent
We can skip intermediate paths in this project
 by going to distances of twins at  large
 enough z where peculiar velocity corrections are not needed. Then we 
 obtain $H_{0}$  directly.

\end{document}